\documentclass[journal=jctc,manuscript=article]{achemso}
\pdfoutput=1
\usepackage[version=3]{mhchem} 
\usepackage{amsfonts}
\usepackage{physics}
\usepackage{dsfont}
\usepackage{amssymb}
\usepackage{bbold}
\usepackage{physics}
\usepackage{geometry}
\usepackage{soul}

\newcommand{\pder}[2]{\frac{\partial #1}{\partial #2}}
\newcommand{\prts}[1]{\left(#1\right)}
\newcommand{\dqq}[1]{\frac{\partial^2 #1}{\partial q^2}}
\def\t{\text}
\def\l{\left}
\def\r{\right}
\def\id{\text{\(\mathds{1}\)}}
\def\aq{{\alpha q}}
\def\bq{{{\beta q}}}
\def\bqp{{\beta q'}}
\def\piaq{\Pi_{\alpha q}}
\def\pibq{\Pi_{\beta q}}
\def\pibqp{\Pi_{\beta q'}}
\def\fullint{\int_{-\infty}^{+\infty}}

\author{Gustavo J. R. Aroeira}
\author{Kyle T. Kairys}
\author{Raphael F. Ribeiro}
\email{raphael.ribeiro@emory.edu}
\affiliation[Emory University]
{Department of Chemistry and Cherry Emerson Center for Scientific Computation, Emory University, Atlanta, Georgia, United States of America}

\title{Coherent transient exciton transport in disordered polaritonic wires}

\SectionNumbersOn

\begin{document}

\begin{abstract}
Excitation energy transport can be significantly enhanced by strong light-matter interactions. In the present work, we explore intriguing features of coherent transient exciton wave packet dynamics on a lossless disordered polaritonic wire. Our main results can be understood in terms of the effective exciton group velocity, a new quantity we obtain from the polariton dispersion. Under weak and moderate disorder, we find that the early wave packet spread velocity is controlled by the overlap of the initial exciton momentum distribution and its effective group velocity. Conversely, when disorder is stronger, the initial state is nearly irrelevant, and red-shifted cavities support excitons with greater mobility. Our findings provide guiding principles for optimizing ultrafast coherent exciton transport based on the magnitude of disorder and the polariton dispersion. The presented perspectives may be valuable for understanding and designing new polaritonic platforms for enhanced exciton energy transport.
\end{abstract}

\clearpage
\listoffigures
\tableofcontents
\newpage

\section*{Introduction}
\par The strong light-matter interaction regime is achieved when the coupling strength between light and matter overcomes dephasing and dissipative phenomena acting on each subsystem. This can be accomplished, for example, with a molecular ensemble with a narrow linewidth bright transition near resonance with an optical microcavity composed of two parallel mirrors with high-reflectivity\cite{kavokin2017microcavities, Yu2018, Tibben2023}. In this scenario, the field confinement and low mode volumes allow light and matter to exchange energy (quasi)reversibly. Several recent studies have shown that strong light-matter coupling can be harnessed to control energy \cite{Coles2014Polariton-mediatedMicrocavity,Zhong2016Non-RadiativeStates,Zhong2017EnergyMolecules,Georgiou2017, Lerario2017High-speedPolaritons, Myers2018Polariton-enhancedTransport, Xiang2020IntermolecularCoupling,Hou2020UltralongRangePropagation,Wang2021Polariton-assistedHeterojunctions,Wei2021,Guo2022BoostingSubstrate, Pandya2022TuningDelocalization, Xu2022UltrafastInteractions, Nosrati2023, Balasubrahmaniyam2023FromExcitations} and charge\cite{Orgiu2015ConductivityField,Krainova2020PolaronRegime,Nagarajan2020ConductivityCoupling,Bhatt2021EnhancedCoupling, Liu2022PhotocurrentExcitonpolaritons} transport in disordered materials. These effects are attributed to the formation of polaritons, i.e., hybrid light-matter states with intermediate properties between purely material or photonic. For example, polariton delocalization \cite{agranovich2003cavity, Shi2014, basko2000electronic, Du2018TheoryTransfer} is often invoked to explain the properties of energy transport in the strong coupling regime\cite{Tibben2023}. 

\par Unlike bare excitons, which tend to show weak delocalization and inefficient energy transfer in disordered media, polaritons show much greater diversity in wave function delocalization \cite{agranovich2003cavity, litinskaya2006loss, michetti2005polariton, suyabatmaz2023vibrational, Engelhardt2023} and transport phenomena \cite{Myers2018Polariton-enhancedTransport, Xu2022UltrafastInteractions, Balasubrahmaniyam2023FromExcitations}. For instance, polariton transport imaging has revealed ultrafast ballistic propagation in perovskite microcavities\cite{Xu2022UltrafastInteractions} and surface-bound polaritons\cite{Hou2020UltralongRangePropagation,Balasubrahmaniyam2023FromExcitations}, with spread velocities spanning several orders of magnitude. The effects of disorder on polariton transport have also received significant attention as the potential source of the slower-than-expected polariton wave packet propagation reported by several groups \cite{Pandya2022TuningDelocalization, Balasubrahmaniyam2023FromExcitations, Xu2022UltrafastInteractions}. Indeed, theoretical investigations suggest that dynamic and static disorder inhibit polariton wave packet propagation by effectively reducing the propagation velocity \cite{Xu2022UltrafastInteractions, Sokolovskii2023, Aroeira2023}. Interestingly, recent theoretical investigations of dipolar exciton propagation in finite one-dimensional systems suggest that under strong disorder, a disorder-enhanced transport regime emerges where coherent exciton propagation benefits from an increase in the static fluctuations of matter excitation energies \cite{Allard2022Disorder-enhancedPhotons, Aroeira2023, Engelhardt2023}. 

\par Our recent work on coherent transport in polaritonic wires \cite{Aroeira2023} thoroughly examined the requirements for convergence of exciton transport simulations with respect to model parameters. We showed that multiple (on and off-resonant) electromagnetic mode \cite{Aroeira2023} played a key role in the exciton dynamics, and demonstrated the potential to control transient ballistic and diffusive exciton transport and Anderson localization under strong light-matter coupling. Here, we focus on the transient early dynamics of exciton wave packets propagating on a lossless polaritonic wire. In particular, we present numerical simulations and a detailed theoretical analysis of coherent polariton-mediated exciton transport in the ballistic regime. Our results and mathematical analysis reveal several surprising aspects of polariton-assisted coherent exciton transport, including a striking difference between the effect of disorder on ultrafast coherent exciton propagation in free space \cite{cui2023} and in a polaritonic medium. Furthermore, we show that an effective exciton group velocity may be defined that allows a qualitative understanding of our numerical simulations even in a moderately disordered scenario.

\par This article is organized as follows: in Section 2, we describe the theory and method employed in this work. Section 3 contains our main numerical results and theoretical analysis, while Section 4 provides conclusions and a summary of this work.

\section*{Theory and Computation}
\par
\begin{figure*}
    \centering
    \includegraphics[width=\textwidth]{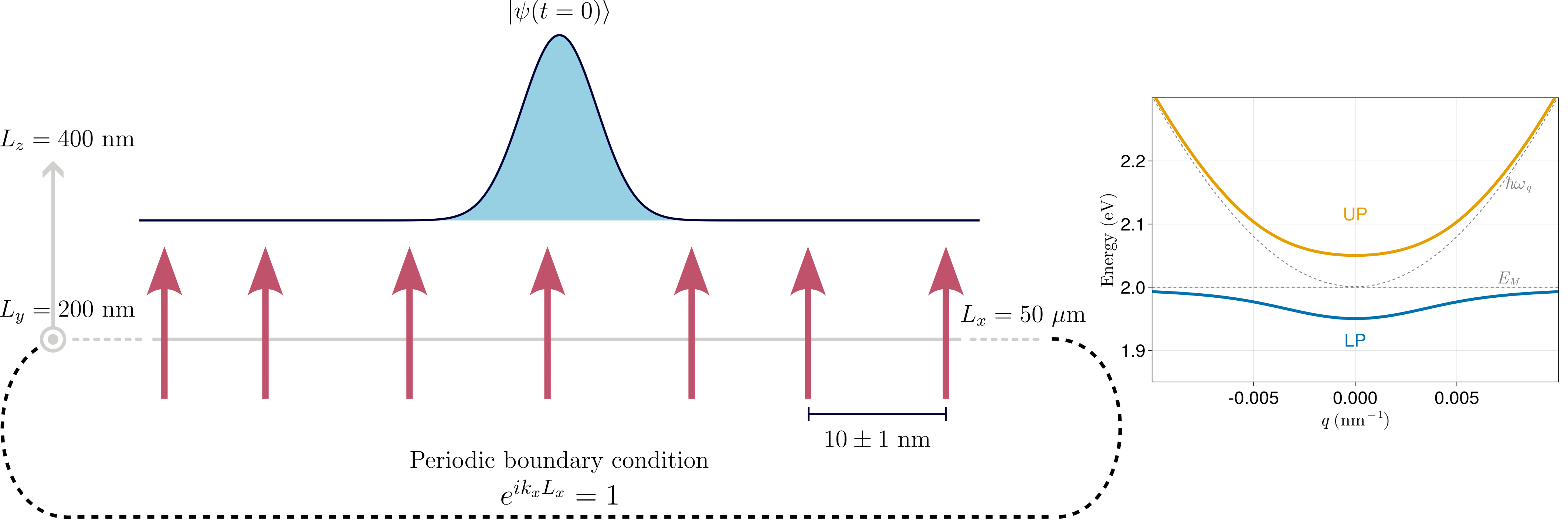}
    \caption{Illustration of the polariton wire model. Dipoles representing matter are non-interacting two-level systems aligned with the z-axis. Distances between sites are sampled from a normal distribution with an average and standard deviation of 10 and 1 nm, respectively. Likewise, excitation energies for each dipole are sampled from a normal distribution using an average of $E_M$ and standard deviation of $\sigma_M$. These parameters control the detuning and static disorder of the system. Radiation states inside the cavity are constructed using 1001 modes of the lowest energy band, with a minimum photon energy of 2.00 eV. On the right side the dispersion for an ordered system ($\sigma_M = 0$) is shown. Continuous lines correspond to UP (orange) and LP (blue) branches and dashed lines represent the bare matter ($E_M$) and empty cavity ($\hbar \omega_q$) dispersions.}
    \label{fig:system}
\end{figure*}

 The polaritonic wire model employed here consists of a linear chain of dipoles representing matter (e.g., atoms, quantum wells, or molecules with negligible vibronic coupling) coupled to photon modes of a lossless cuboid optical microcavity of lengths $L_x$, $L_y$, and $L_z$ as depicted in Figure \ref{fig:system}. Each dipole is a two-level system with excitation energy given by $\hat{H}_M |n;0\rangle = E_n |n;0\rangle$, where $|n;0\rangle$ represents a state where the $n$-th dipole is in its excited state, while all other dipoles and cavity modes are in their ground states. The excitation energy of each dipole $E_n$ is sampled from a normal distribution with average $E_M$ and standard deviation $\sigma_M$. Different detuning and static disorder strengths are accessed by adjusting these parameters. The spatial distribution of dipoles is also sampled from a normal distribution, but in this case, the average and standard deviation are fixed at 10 and 1 nm, respectively. The large intersite separation allows us to disregard direct interaction between these dipoles since direct energy transfer via FRET would occur at a much larger times scale than probed here. Furthermore, we also impose the same orientation for all dipoles, such that they can only interact with transverse electric (TE) polarized photon modes, and we need not consider the transverse magnetic polarization.

\par Imposing vanishing electric field along the $y$ and $z$ directions and periodic boundary conditions along the long-axis $x$ implies
\begin{align}
    \mathbf{k} = \left(\frac{2\pi m_x}{L_x},\; \frac{\pi n_y}{L_y},\; \frac{\pi n_z}{L_z} \right) \,,
\end{align}
where $m_x \in \mathbb{Z}$ and $n_y, n_z \in \mathbb{N}_{>0}$. Throughout this work, we employ a geometry where $L_x = 50$ $\mu m$, $L_y = 0.2$ $\mu m$, and $L_z = 0.4$ $\mu m$, such that the energy gap between adjacent bands is large enough ($> 0.5$ eV) that we restrict our analysis to the lowest energy band ($n_y = 1$ and $n_x = 1$). By defining $q = k_x = \frac{2\pi m_x}{L_x}$ and $q_0 = \sqrt{k_x^2 + k_y^2} = \sqrt{\left(\frac{\pi}{L_x}\right)^2 + \left(\frac{\pi}{L_y} \right)^2}$, it follows we can uniquely identify each photon mode using its value of $q$. The energy of each mode is
\begin{align}
    \hbar \omega_q = \frac{\hbar c}{\sqrt{\epsilon}} \sqrt{q^2 + q_0^2}\,, \label{photenergy}
\end{align}
where $\hbar$ is the reduced Planck constant, $c$ is the speed of light, and $\epsilon$ is the relative permittivity of the intracavity medium. We use $\epsilon = 3$ as a suitable parameter for organic microcavities. From Equation \eqref{photenergy}, the minimum photon energy supported in the cavity is $\hbar c q_0/\sqrt{\epsilon} = 2.00$ eV. 

\par The non-interacting part of the light-matter Hamiltonian is
\begin{align}
    \hat{H}_0 = \sum_n E_n \hat{b}^\dagger_n \hat{b}_n + \sum_q \hbar \omega_q \hat{a}^\dagger_q \hat{a}_q \,,
\end{align}
where $\hat{b}^\dagger_n$ and $\hat{a}^\dagger_n$ are ladder operators creating a dipolar excitation at the $n$-th site and a photon in the $q$ mode, respectively. We employ the Coulomb gauge in the rotating wave approximation while neglecting the diamagnetic contribution. It follows the interacting part of the Hamiltonian can be expressed as
\begin{align}
    \hat{H}_\text{int} = \sum_{n=1}^{N_M}\sum_q \frac{-i\Omega_R}{2}\sqrt{\frac{E_n}{N_M\hbar\omega_q}}\left(e^{iqx_n}\hat{b}_n^\dagger \hat{a}_q - e^{-iqx_n}\hat{a}_q^\dagger \hat{b}_n\right)\,,
\end{align}
where $N_M$ is the total number of sites and $x_n$ is the position of the $n$-th dipole along $x$. The parameter $\Omega_R$ (Rabi splitting) is related to the transition dipole moment of each molecule ($\vec{\mu_d}$) by 
\begin{align}
    \Omega_R = \vec{\mu_d}\cdot \vu{z} \sqrt{\frac{\hbar\omega_0 N_M}{2\epsilon L_x L_y L_z}}\,.
\end{align}

\par To make our study computationally tractable, we truncate the model in the number of molecules and photon modes. Following the thorough analysis in our previous work \cite{Aroeira2023}, we set $N_M = 5000$ to minimize finite-size effects in the sub-picosecond region. Similarly, we include 1001 cavity modes ($-500 \leq m_x \leq 500$), which span an energy range of 7.43 eV, well above the necessary for convergent results. 

\par In all simulations presented here, the initial state is a Gaussian exciton wave packet with zero photonic content. This initial state can be represented in the uncoupled basis as
\begin{align}
    \ket{\psi(0)} = \frac{1}{Z}\sum_n \exp\left[-\frac{(x_n-\frac{1}{2}L_x)^2}{4\sigma_x^2} +i\bar{q}_0x_n\right] \ket{n;0}\,, \label{gaussianwvp}
\end{align}
where $Z$ is a normalization constant,  $\sigma_x$ is the initial spread of the wave packet, and $\bar{q}_0$ is the average exciton momentum along $x$. From now on, we set $\bar{q}_0 = 0$ unless otherwise noted. Note that $\sigma_x$ is the standard deviation of the probability distribution $P(n) = |\braket{n;0}{\psi(0)}|^2$. The reciprocal space distribution $P(q)$, obtained from the Fourier change of basis in \eqref{gaussianwvp}, has standard deviation $\sigma_q$ related to $\sigma_x$ via 
\begin{align}
    \sigma_x \sigma_q = \frac{1}{2}\;. \label{uncert}
\end{align}

We note it may be possible to experimentally prepare exciton wave packets similar to those examined by us with surface plasmon-exciton polariton systems. In these, the molecular ensemble can be directly excited, and the incidence angle and excitation spot size may be used to control the resulting initial state (e.g., as done in ref. \citenum{Vasa2013}).

\par In all simulations, the Fock space is truncated to include only states with one excited dipole and no photons ($\ket{n;0}$) or one photon and no dipolar excitations ($\ket{0;q}$). The dynamics generated by $\hat{H} = \hat{H}_0 + \hat{H}_{\text{int}}$ and the initial state $\ket{\psi(0)}$ is in fact constrained to the single excitation subspace. Thus, our results are relevant to coherent dipolar dynamics when nonlinearities can be ignored (e.g., due to a small density of excitons). The time-evolved wave packet is directly obtained from $\ket{\psi(t)} = e^{-i\hat{H}t/\hbar}\ket{\psi(0)}$ using the eigenvalues and eigenvectors of $\hat{H}$  within the one-excitation manifold ($\ket{n;0} \oplus \ket{0;q}$). 

\par Our computational study follows the transient evolution of exciton wave packets starting from the well-localized purely excitonic state in Equation \eqref{gaussianwvp}. As a metric for the exciton spread, we compute the root mean square displacement of the dipolar component of the wave packet, defined here as
\begin{align}
    \text{RMSD}(t) &= \left[\frac{1}{P_\text{M}(t)}\sum_{n=1}^{N_M} |\bra{n;0}\ket{\psi(t)}|^2(x_n - x_0)^2\right]^{1/2}\,, \label{x2}\\
    P_\text{M}(t) &= \sum_n^{N_M} |\bra{n;0}\ket{\psi(t)}|^2 \label{PM}\,.
\end{align}
where the renormalization factor $P_M(t)$ is the time-dependent probability of finding any excited dipole and $x_0$ is the average exciton position at $t=0$ (which in this work is always $\frac{L_x}{2} = 25\;\mu$m). As an alternative and complementary metric of exciton mobility, we define the (matter normalized) migration probability $\chi(t)$, which measures the conditional probability of finding an excited dipole outside the region where the wave packet was initially localized. By choosing symmetric boundaries $n_\text{min} \leq n \leq n_\text{max}$ such that there is at least 99\% chance ($\frac{L_x}{2} -3\sigma_x < x < \frac{L_x}{2} + 3\sigma_x$) that the exciton at time $t = 0$ lies within this region, we compute the migration probability as
\begin{align}
    \chi(t) = 1 - \frac{1}{P_\text{M}(t)}\sum_{n = n_\text{min}}^{n_\text{max}} |\bra{n;0}\ket{\psi(t)}|^2 \label{escape} \,.
\end{align}

\par All results including disorder are averages from 100 realizations. The code used in all simulations is available in our prototype package \textsc{PolaritonicSystems.jl} \cite{psys}. Random variables were generated using the \textsc{Distributions.jl} package\cite{Besancon2019} and the \textsc{Makie.jl} plotting ecosystem \cite{makie} was used for data visualization. 

\section*{Results and Discussion}
\subsection*{Polariton-mediated exciton wave packet propagation }
\par
\begin{figure*}
    \centering
    \includegraphics[width=\textwidth]{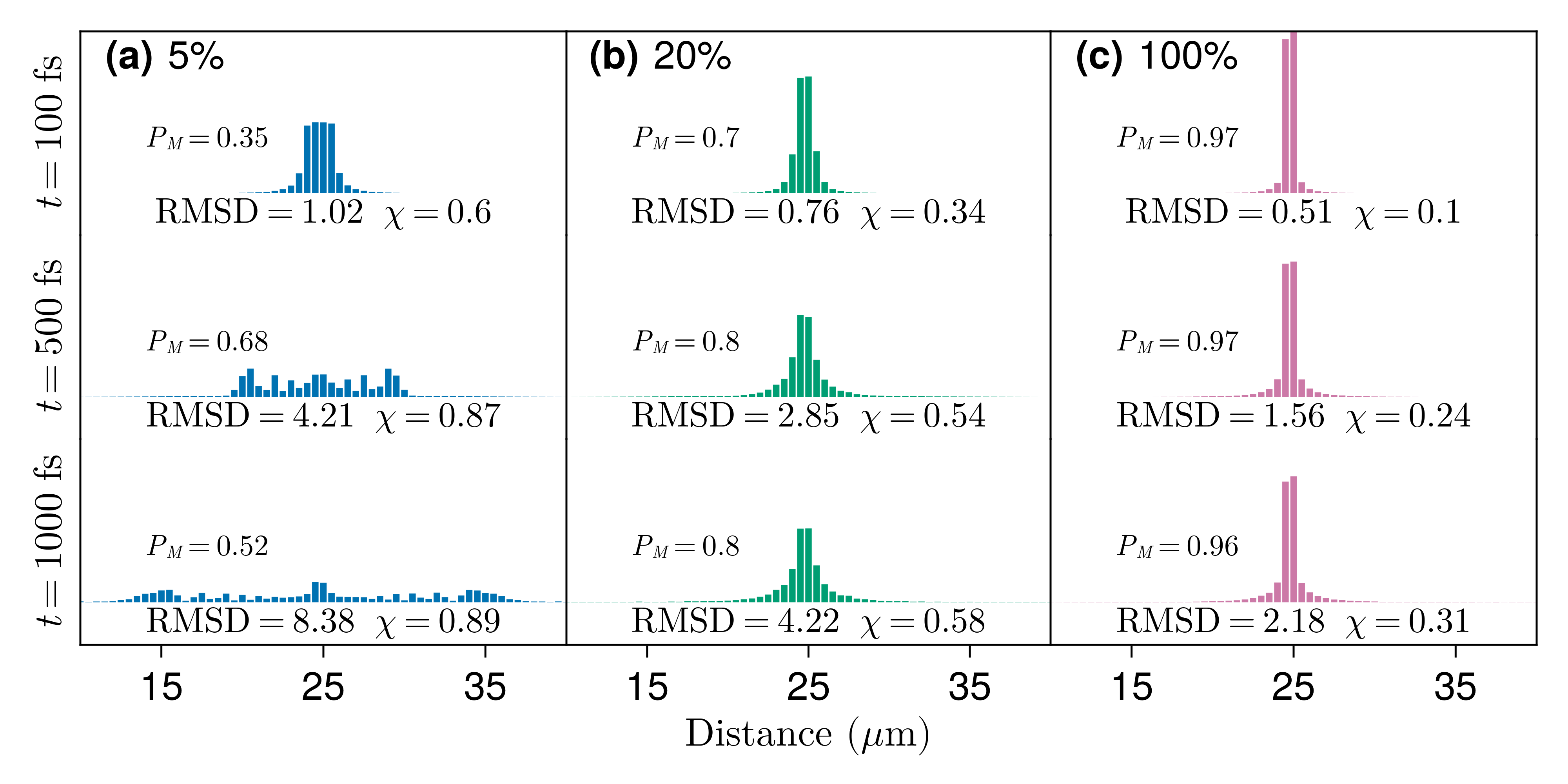}
    \caption{Average exciton wave packet profiles at different time delays and relative disorder strength ($\sigma_M/\Omega_R$) of 5\%, 20\%, and 100\% for (a), (b) and (c), respectively. Probabilities are grouped in bins containing 50 dipoles spanning 0.5 $\mu m$.  $P_M$, RMSD, and  $\chi$ are defined in Equations \eqref{x2},\eqref{PM}, and \eqref{escape}, respectively. In all cases, $\Omega_R = 0.1$ eV and 
 $\sigma_x = 120$ nm.}
    \label{fig:wvp}
\end{figure*}

 Selected wave packet snapshots are given in Figure \ref{fig:wvp}, along with the corresponding time-dependent exciton RMSD, $P_M$, and migration probability $\chi$. The main effect of static disorder can be observed in these examples. From Figures \ref{fig:wvp}a to c, as disorder is increased, the wave packet mobility is significantly reduced, and its spread is strongly suppressed. Simultaneously, we find the photonic content and its time-dependent fluctuations monotonically decrease and become small under strong disorder (e.g., $P_M(t)$ is relatively stable around 0.97 when $\sigma_M/\Omega_R =100\%$). In contrast, photon content fluctuations are large under weak disorder ($\sigma_M/\Omega_R \rightarrow 0$). The effect of disorder on photon content fluctuations may be directly understood from  Rabi oscillations, which occur unperturbed at weak disorder while being strongly damped as $\sigma_M$ approaches  $\Omega_R$ \cite{Aroeira2023}. These observations illustrate how the oscillatory energy exchange between radiation and matter leads to enhanced coherent exciton transport. Since we are examining very large values of disorder ($\sigma_M/\Omega_R \geq 100\%$), we include in our SI (section 8) an analysis of how the signatures of strong light-matter coupling change under increasingly stronger static disorder.

\par
\begin{figure}
    \centering
    \includegraphics[width=0.45\textwidth]{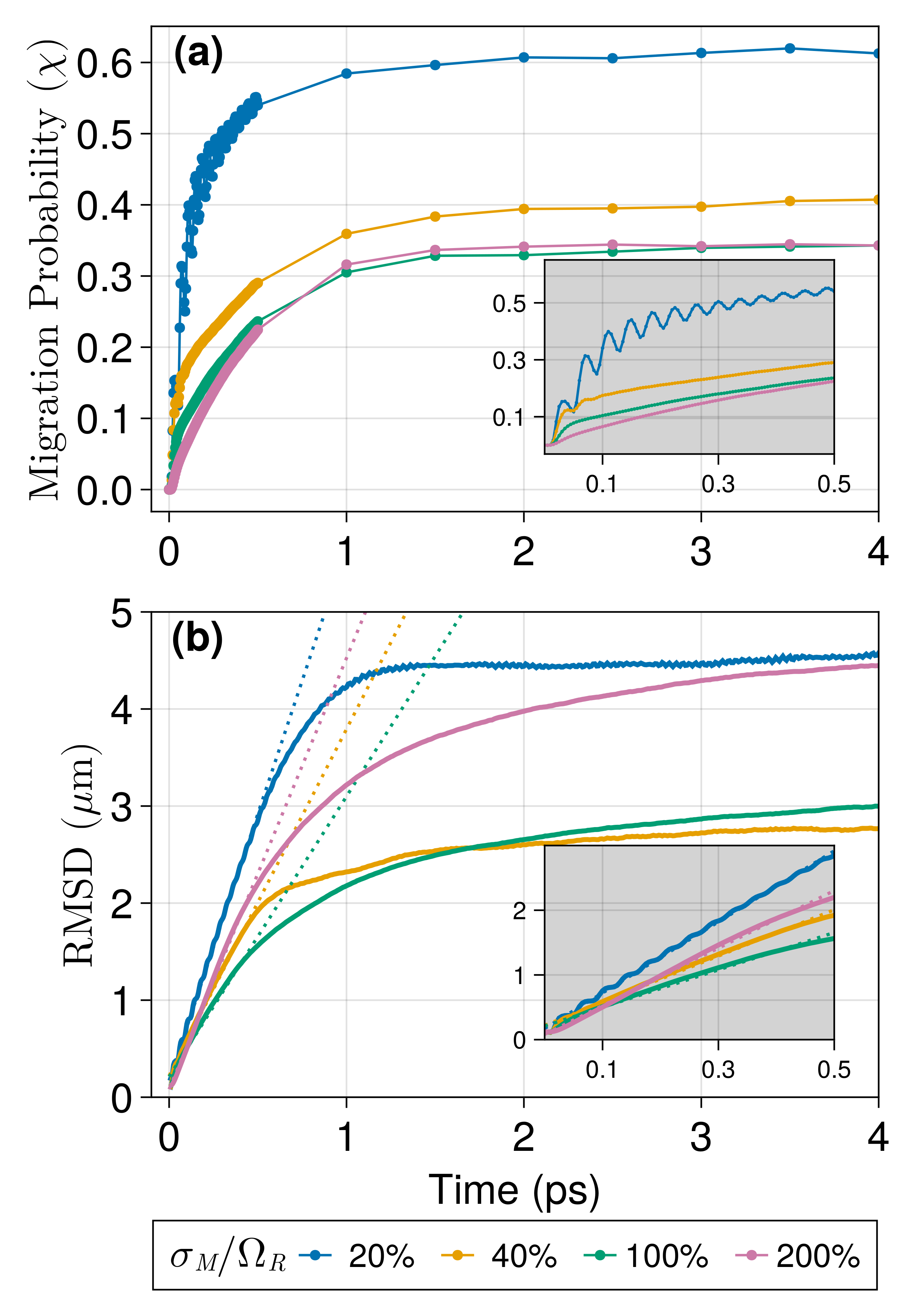}
    \caption{Propagation of exciton wave packets at different disorder strengths measured by \textbf{(a)} migration probability (Equation \eqref{escape}) and \textbf{(b)} RMSD (Equation \eqref{x2}).  The dotted lines in \textbf{(b)} are linear fits of the early propagation (< 500 fs) from which slopes are used to measure the initial ballistic velocity ($v_0$). Insets show a zoomed-in view into the sub 500 fs region. In all cases, $\Omega_R = 0.1$ eV and $\sigma_x = 120$ nm.}
    \label{fig:prop}
\end{figure}

 In Figure \ref{fig:prop}, the average exciton migration probability (Equation \ref{escape}) and RMSD are shown for excitons propagating over 4 ps under different values of relative disorder strength ($\sigma_M/\Omega_R$). The migration probability, seen in Figure \ref{fig:prop}a, increases rapidly before achieving a steady state ($\mathrm{d}\chi(t)/\mathrm{d}t \approx 0$) around 2 ps irrespective of the disorder strength. Conversely, disorder plays a crucial role in the sub-300 fs phase of the dynamics, where we find from the inset that in all considered cases, an increase in $\sigma_M$ leads to a slower initial exciton migration. From the behavior of $\chi(t)$ at large $t$, we find the steady-state exciton migration probability at weak disorder is largely suppressed when $\sigma_M$ is increased. Nevertheless, the strongly disordered cases ($\sigma_M/\Omega_R \geq 1$) suggest that beyond a particular disorder strength, $\chi(t\gg 0)$ becomes approximately independent of $\sigma_M$. Figures S6 and S7 provide $\chi(t)$ for $\Omega_R = 0.2$ eV and $\Omega_R = 0.3$ eV under different levels of disorder and verify the disorder effects on $\chi(t)$ described above are in fact generic.   

\par In the SI Section 2, we show the initial growth of $\chi(t)$ can be approximated in the weak and strong disorder limits by analyzing the disorder-averaged properties of $|\braket{0;n}{\psi(t)}|^2$. This leads to the conclusion that $\partial\chi(t)/\partial t \rightarrow 0$ as $t\rightarrow 0^+$, so that the quantity $G=(1/2)\partial^2 \chi(t)/\partial t^2, ~t\rightarrow 0^+$ controls the early growth of $\chi(t)$. In the weak and strong disorder limits, $G$ satisfies, respectively
\begin{align}
   & G^W  \approx \frac{1}{2N_{\mathcal{I}}} \sum_{A, B\neq A} \sum_{n\in \mathcal{I}}  |A_n|^2 |B_{n}|^2 (\omega_A-\omega_B)^2, \label{eq:G_nd}\\
   & G^{S}  \approx  \frac{1}{2}\sum_{A, B\neq A}  \sum_{n\in \mathcal{I}} \overline{|A_n|^2 |B_n|^2} |c_n|^2 (\omega_A-\omega_B)^2, \label{eq:Gd}
\end{align}
where $A$ and $B$ are eigenstates, $A_n$ and $B_n$ the probability amplitude to detect an exciton at the $n$th dipole when the system is in the $A$ and $B$ eigenstates, respectively, $\mathcal{I} = [n_{\text{min}},n_{\text{max}}]$ (see Equation \eqref{escape} and accompanying description), $N_{\mathcal{I}}$ is the number of sites in $\mathcal{I}$, $c_n$ is the $n$th exciton amplitude in the initial wave packet, and $\overline{f}$ is the disorder average of quantity $f$. Both approximations to $G$ in the weakly and strongly disordered regimes suggest that an ultrafast increase in the exciton migration probability depends on the existence of eigenstates with large energy differences and significant contributions from the dipoles comprising the initial wave packet. 
\par From Equation \eqref{eq:Gd}, we infer (i) the steep increase of $\chi(t)$ at early times is enhanced by raising $\Omega_R$ at fixed energetic disorder (as the energy difference between polariton modes formed from near-resonant photon and excitons increase with $\Omega_R$), (ii) increasing disorder with fixed $\Omega_R$ leads to initial slower growth of $\chi\left(t\rightarrow 0^+\right)$, due to the greater tendency of localization of the $n$-th exciton into strongly localized eigenmodes. Importantly, while the summand of Eq. \eqref{eq:G_nd} has the same form as that of Equation \eqref{eq:Gd}, the former has several more non-negligible contributions than the latter ($N_{\mathcal{I}}^2$ in the approximation given in Equation \eqref{eq:G_nd}), and therefore, a much steeper early increase occurs in $\chi(t)$ under weak disorder.  
\par In summary, as measured by $\chi(t)$, indeed, disorder slows down the ability of excitons to migrate at very early times, and increasing the Rabi splitting leads to faster initial migration probability for dipoles in photonic wires. Similar considerations can be made on the asymptotic ($t\rightarrow \infty)$ behavior of $\chi(t)$: disorder averaging and the lack of correlation between the excitation energy at distinct sites suppresses cross-terms in the strong disorder limit relative to weak, and therefore, a reduced $\sigma_M$ favors a larger steady-state value of $\chi(t)$. Additionally, Figure \ref{fig:prop}a suggests a steady-state time for $\chi(t)$ that is almost independent of disorder. This contrasts with the RMSD behavior shown in Figure 3b, implying that the approximately disorder-independent $\chi(t)$ steady-state times correspond to a feature specific to the non-standard observable $\chi(t)$. For the rest of this manuscript, we characterize the transport velocity using the RMSD, therefore we leave a detailed analysis of $\chi(t)$ for future studies.

\par The RMSD measure reported in Figure \ref{fig:prop}b, also indicates the excitonic propagation is fastest in the fs time scale. Both  $\chi(t\rightarrow \infty)$ and RMSD($t\rightarrow\infty$) drop when the $\sigma_M/\Omega_R$ is increased from 20\% to 40\%. However, comparing the 40\% and 100\% relative disorder strength cases in Figure \ref{fig:prop}b, we note the emergence of a disorder-enhanced transport regime (DET) as reported in previous studies of dipole chains under strong light-matter interactions \cite{Chavez2021Disorder-EnhancedCavities,Allard2022Disorder-enhancedPhotons, Aroeira2023, Engelhardt2023}. This DET regime clearly leaves no signature in $\chi(t)$ (Figure \ref{fig:prop}a), but may be understood based on earlier studies of exciton transport in a polaritonic wire. In this regime, weakly coupled states are expected to be exponentially localized but can have slowly decreasing extended tails\cite{Allard2022Disorder-enhancedPhotons,Engelhardt2023}. These tails carry a small probability away from the initial excitation spot and contribute to the greater RMSD values reported with increasing $\sigma_M/\Omega_R$. Conversely, the extended tails associated with DET leave no signature in $\chi(t)$, as this quantity only tracks the probability in the bulk of the wave packet. These points are corroborated by Figures S3-S5 of the SI, where we show the decay profile of wave packets and the emergence of the extended tails responsible for DET.

We reinforce that the asymptotic behavior seen in both Figures \ref{fig:prop}a and b can be attributed to Anderson localization and not to finite-size artifacts. To demonstrate that, we show in Figure S1 the shape of the wave packets at long propagation times whereas Figure S2 reveals that, under sufficiently strong disorder, the exciton probability at $x_0 \pm L_x/2$ (i.e., where periodic boundary conditions are enforced) remains negligible throughout the simulation.

In Figure \ref{fig:prop}b, dotted lines represent linear fits obtained from the first 500 fs of simulation. This initial linear behavior (minimum coefficient of determination $R^2 = 0.98$) characterizes the excitonic ballistic spread. In the next sections, we will use this value as a measure of the initial exciton spread velocity ($v_0$). Our choice of a 500 fs time interval aimed to average out complicated features at very early times and to avoid localization effects observed around 1 ps. We checked that all reported qualitative trends are unaffected by a choice of initial time interval that satisfies the described conditions (see Figures S10-S12 for a comparison of $v_0$ obtained using different initial time intervals).

The existence of a short-lived ballistic regime even under strong disorder ($\sigma_M/\Omega_R \geq  1)$ is a generic feature of sufficiently narrow wave packets. It follows from the fact that the $n$-th exciton probability at early times $t \rightarrow 0$ is given by $P_n(t) \approx P_n(0) + a_n t^2$, where $a_n$ is a site-dependent constant\cite{Moix2013}. Therefore, $\text{RMSD}^2(t)$ at sufficiently small $t$ is equal to RMSD$^2$(0) + $\sum_{n} (x_n-x_0)^2 a_n t^2$. Hence, there is a short time frame where RMSD(t) = $\sqrt{\text{RMSD}^2(0) + c t^2}$, where $c = \sum_{n} (x_n-x_0)^2 a_n$. If RMSD(0) is small enough, then there exists a short time interval where $\text{RMSD}(t)$ is proportional to $t$ regardless of the amount of disorder.

To conclude this subsection, we note the polariton-mediated ultrafast exciton transport described here shows intriguing differences relative to bare exciton transport analyzed in a recent study by Cui et al.\cite{cui2023}. Their work showed that transient ultrafast energy transfer mediated by direct short-range interactions benefits from the existence of static disorder, leading to faster transport (relative to a perfectly ordered system) in the femtosecond timescale. Cui et al. ascribe their observation of transient disorder enhancement of transport to the suppression of destructive interference induced by the heterogeneity of the matter excitation energies \cite{cui2023}. Here, we find the opposite feature: disorder always reduces the initial transport velocity. Even in the DET regime, Figure \ref{fig:prop}b shows a narrow window at early times where transport is subdiffusive \cite{Aroeira2023}. This contrast points towards a fundamental difference in how static disorder affects direct and polariton-mediated coherent exciton transport.

\subsection*{Ballistic Exciton Transport}
\par
\begin{figure*}
    \centering
    \includegraphics[width=0.8\textwidth]{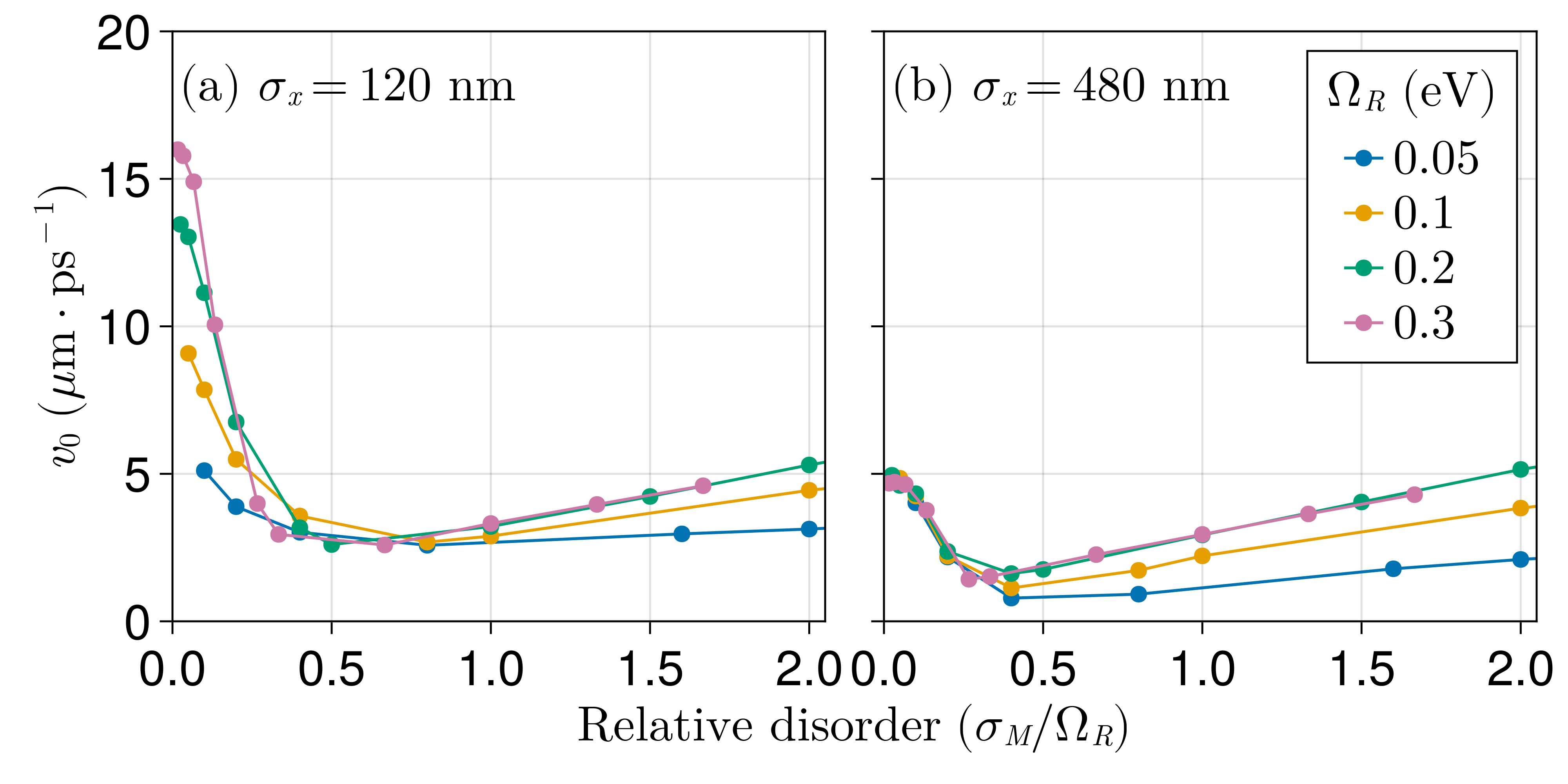}
    \caption{Initial ballistic velocity ($v_0$) for various wave packets with a \textbf{(a)} narrow and \textbf{(b)} broad initial spread values ($\sigma_x$, see Equation \eqref{gaussianwvp}).  $v_0$ was computed as the slope of a linear fit of RMSD values in the initial 500 fs of simulation (see Figure \ref{fig:prop}).}
    \label{fig:v0}
\end{figure*}

 In Figure \ref{fig:v0}, the initial exciton spread velocity ($v_0$) is shown as a function of relative disorder $\sigma_M/\Omega_R$. We examine $v_0$ obtained for systems with variable collective light-matter interaction strength $\Omega_R$ (with fixed relative disorder $\sigma_M/\Omega_R$) and two selected initial wave packet sizes ($\sigma_x$ in Equation \eqref{gaussianwvp}). In both cases, we observe an initial steep decay of $v_0$ with increasing $\sigma_M/\Omega_R$ which is followed by a plateau until the DET regime is reached at $\sigma_M/\Omega_R \approx 1$. However, a salient difference in the variation of $v_0$ with $\Omega_R$ at low disorder is observed between the narrow ($\sigma_x = 120$ nm) and the broader ($\sigma_x = 480$ nm) wave packets in Figures \ref{fig:v0}a and b, respectively. This difference vanishes quickly when disorder is increased, demonstrating the initial state preparation is less important to the dynamics under strong disorder. Nevertheless, the distinct $\Omega_R$ dependence of $v_0$ is observable in a sizable range of disorder strengths (0$\sim$30\%), thereby warranting a mechanistic explanation. We pursue that by analyzing below the (excitonic) spread velocity of the wave packet in the absence of disorder. 

\par The detailed mathematical treatment of the spread velocity, which we summarize below, is provided in Section 1 of the SI. We first emphasize that the treatment of the ballistic transport regime is unconventional even in the zero-disorder case because (a) our initial wave packets range from strongly localized ($\sigma_x = 120 ~\text{nm})$ to moderately delocalized $(\sigma_x = 480~\text{nm})$ in real space, (b) the wave packet has LP and UP components, and (c) the polariton dispersion is not quadratic. These features imply the basic treatment of Gaussian wave packet transport in a quadratic medium, generally valid for sufficiently narrow wave packets in $q$-space, is inapplicable \cite{elmore1985physics}. With these considerations, we show (see SI) that the dominant contribution to the exciton transport velocity $v_0$ is given by
\begin{align}
    {v}^2_0 \approx \sum_q P(q) \left[ \left(v_{\text{LP}q}^\text{eff}\right)^2  +  \left( v_{\text{UP}q}^\text{eff} \right)^2 \right] \;, \label{v02}
\end{align}
where $P(q)$ is the $t=0$ exciton probability distribution in $q$-space. From Equation \eqref{uncert}, the width $\sigma_q$ of $P(q)$ is inversely proportional to the real-space width of the initial wave packet $\sigma_x$. The effective group velocity $v_{\alpha q}^\text{eff}$ (where $\alpha$ is UP or LP) is defined as
\begin{align}
    v_{\alpha q}^\text{eff} = \Pi_{\alpha q} v_g^{\alpha q} = \Pi_{\alpha q} \frac{\partial \omega_{\alpha q}}{\partial q} \label{vgexc}\; ,
\end{align}
where $\Pi_{\alpha q}$ is the total exciton content of the polariton mode $\alpha$ with wave number $q$, and $\frac{\partial \omega_{\alpha q}}{\partial q}$ is the (conventional) group velocity $v_g^{\alpha q}$ of the same mode. The polaritonic group velocity weighted by the corresponding exciton content $\Pi_{\alpha q}$ yields the effective exciton group velocity $v_{\alpha q}^\text{eff}$ of mode $\ket{\alpha q}$. The total matter content $\Pi_{\alpha q}$ plays a key role because even though high energy regions of the UP branch yield the largest $v_g^{\alpha q}$, their small $\Pi_{\alpha q}$ results in negligible $v_{\alpha q}^\text{eff}$ values. From Equation \eqref{v02} one can see that $v_0$ is controlled by the effective group velocity $v_{\alpha q}^\text{eff}$ weighted by the exciton $P(q)$ distribution. Hence, as we demonstrate below, the different ways $P(q)$ and $v_{\alpha q}^\text{eff}$ overlap explain the varying mobility of differently prepared excitons in weakly and moderately disordered systems.

\par
\begin{figure*}
    \centering
    \includegraphics[width=0.9\textwidth]{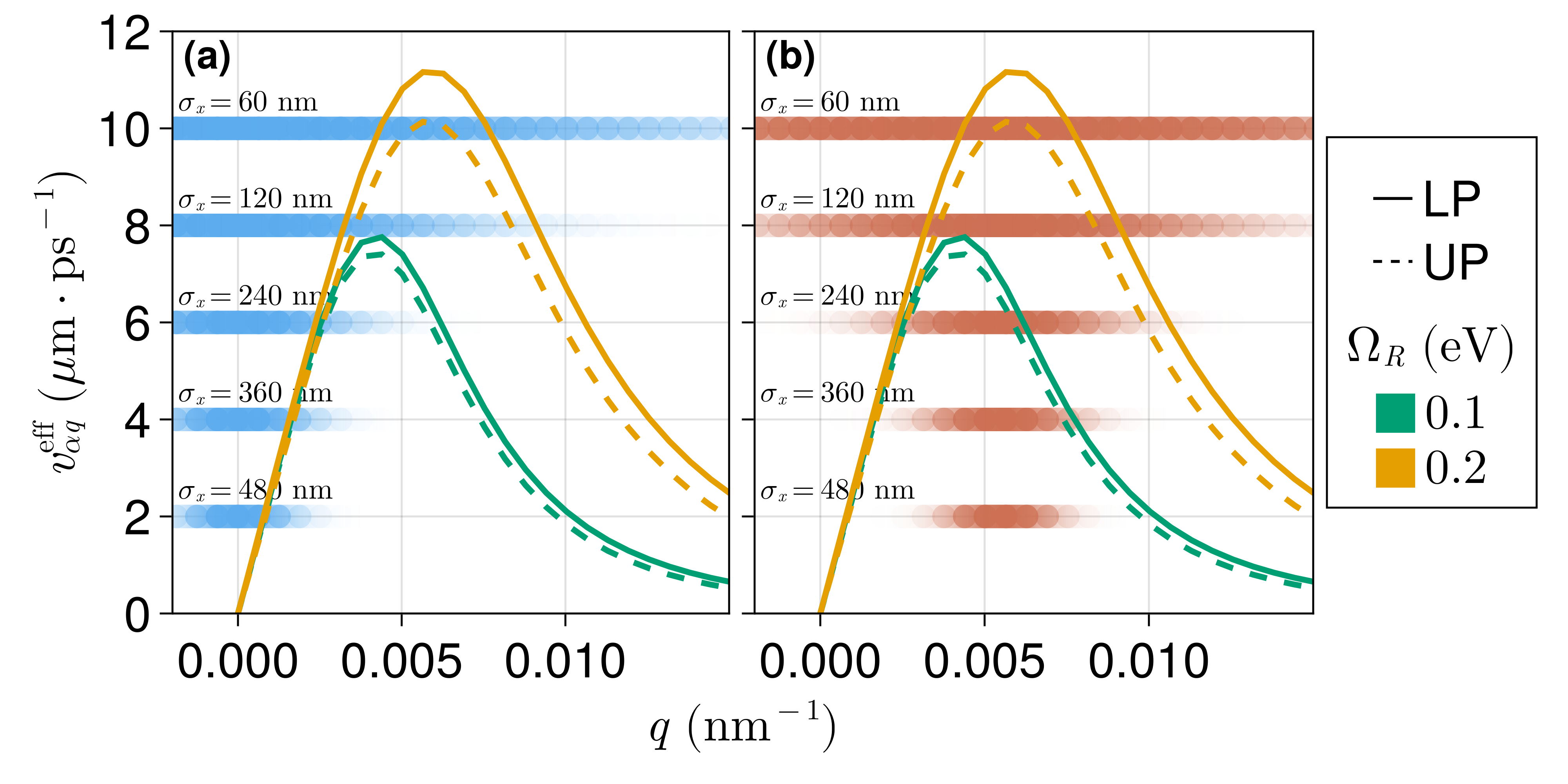}
    \caption{Effective group velocity (Equation \eqref{vgexc}) for different values of $\Omega_R$. Horizontal gradient bars represent the $P(q)$ distribution for different exciton wave packets (see Equation \eqref{gaussianwvp}) with \textbf{(a)} $\bar{q}_0 = 0.0$ and \textbf{(b)} $\bar{q}_0 = 0.0055$ nm$^{-1}$. The overlap between the gradient bars ($P(q)$) and the $v_{\alpha q}^{\text{eff}}$ curves yields the exciton ballistic velocity $v_0$ (Equation \eqref{v02}). See text for more details.}
    \label{fig:vgrabi}
\end{figure*}

 Effective group velocities $v_{\alpha q}^{\text{eff}}$ for systems with $\Omega_R = 0.1$ and $\Omega_R = 0.2$ eV are presented with an overlay of $P(q)$ for several initial exciton wave packets in Figure \ref{fig:vgrabi}. The first relevant observation is the substantial increase of $v_{\alpha q}^{\text{eff}}$ with $\Omega_R$ for both $\alpha = \text{LP}$ and UP with $q > 0.003$ nm$^{-1}$. The Rabi splitting does not affect the effective group velocity for polariton modes with $q < 0.003$ nm$^{-1}$. This feature explains the Rabi splitting dependence of $v_0$ observed in Figure \ref{fig:v0}b in weakly and moderately disordered systems. Specifically, a broad wave packet in real space (e.g., $\sigma_x = 480$ nm) is narrow in $q$-space and is only non-vanishing at a small interval of $q$ near zero (the wave packet center in $q$ space)  where  $v_{\alpha q}^\text{eff}$ is nearly identical for all values of $\Omega_R$ examined here. Therefore, as seen in Figure \ref{fig:v0}b, sufficiently broad wave packets show no significant dependence on $\Omega_R$ when $\sigma_M/\Omega_R$ is not too large (up to 40$\%$ in Figure \ref{fig:v0}).

\begin{figure}
    \centering
    \includegraphics[width=0.45\textwidth]{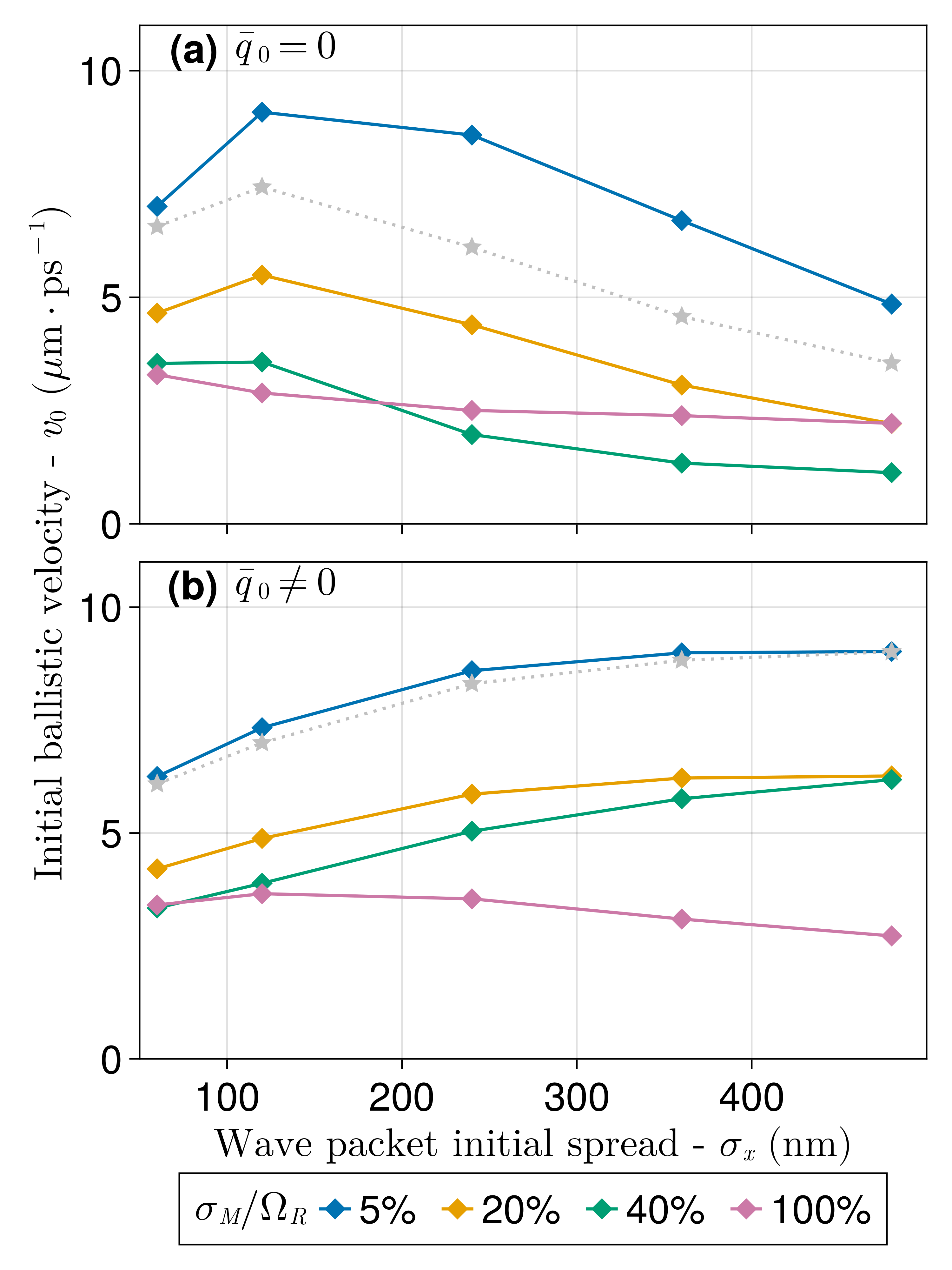}
    \caption{Initial ballistic velocity ($v_0$) dependency on the wave packet initial spread ($\sigma_x$) with: \textbf{(a)} $\bar{q}_0 = 0$ and \textbf{(b)} $\bar{q}_0 = 0.0056$ nm$^{-1}$. Gray dotted curves are values obtained from Equation \eqref{v02}.}
    \label{fig:sigmax}
\end{figure}

In Figure \ref{fig:sigmax}, we present quantitative evidence that Equations \eqref{v02} and \eqref{vgexc} appropriately describe the early-time exciton wave packet propagation rate under small and moderate disorder conditions. In particular, Figure \ref{fig:sigmax} shows $v_0$ vs. $\sigma_x$ at various relative disorder strengths ($\sigma_M/\Omega_R$). As shown in Figure \ref{fig:sigmax}a, in almost every case considered, when the initial state has zero average momentum ($\bar{q}_0 = 0$), an increase in $\sigma_x$ results in a slower propagation. This generic behavior at weak and moderate disorder can be readily understood from Figures \ref{fig:vgrabi}a and b. As $\sigma_x$ broadens, the width of the $P(q)$ distribution (centered at $\bar{q}_0 = 0$) is reduced, and $v_0$ decreases due to the increasing dominance of contributions with small effective group velocities in Equation \eqref{v02}.

 \par The gray dotted curves in Figure \ref{fig:sigmax} follow from Equation \eqref{v02}. This Equation not only captures the overall qualitative trend at small and moderate disorder, but it also reproduces the local maximum at $\sigma_x = 120$ nm in Figure \ref{fig:sigmax}a. This peak can be explained again based on Figure \ref{fig:vgrabi}: by increasing the wave packet width in $q$-space, polariton components with larger $v_{\alpha q}^{\text{eff}}$ become relevant and $v_0$ (Equation \eqref{v02}) increases, but if $P(q)$ becomes too broad (e.g., $\sigma_x = 60$ nm), polaritons with smaller $v_{\alpha q}^\text{eff}$ (with $q$ greater than the maxima in $v_{\alpha q}^{\text{eff}}$) start to contribute significantly to $v_0$ (at the expense of high exciton group velocity components) leading to an overall reduction in the magnitude of $v_0$. 
 
\par Figure \ref{fig:sigmax}b shows analogous results for an exciton prepared with $\bar{q}_0 = 0.005654$ nm$^{-1}$. Indeed, in this case, broader wave packets display higher mobility than narrow ones as measured by $v_0$. This is expected based on Equations \eqref{v02} and \eqref{vgexc} as here a smaller uncertainty in $q$ (increased $\sigma_x$) localizes $P(q)$ around $q\approx \bar{q}_0 \neq 0$ where the corresponding $v_{\alpha q}^\text{eff}$ values are appreciable (see Figure \ref{fig:vgrabi}b). Overall, these results give a simple prescription to optimize polariton-mediated coherent exciton transport by preparing a sufficiently broad initial state (large $\sigma_x$ and small $\sigma_q$) with average momentum ($\bar{q}_0$) centered at the maximum of $v_{\alpha q}^\text{eff}$. 

\par Note that polariton states no longer have well-defined $q$ values in the presence of nonvanishing disorder, and strictly speaking, the arguments above based on uncertainty relations break down. However, in Figure \ref{fig:sigmax}, this breakdown is only observed when $\sigma_M/\Omega_R = 1.0$, where $v_0$ is nearly independent of $\sigma_x$. Our analysis in terms of uncertainty relations and Equations \eqref{v02} and \eqref{vgexc} is seen to hold qualitatively for $\sigma_M/\Omega_R < 40\%$, indicating that the model presented here can be used for the examination of coherent exciton transport at early times even in systems with moderate disorder.

\par To conclude, we investigate the effect of light-matter detuning $\delta = \hbar\omega_0 - E_M$ on polariton-assisted exciton propagation. This study is motivated by detuning being a simple controllable microcavity parameter \cite{kavokin2017microcavities}, and by previous work which reported greater steady-state exciton migration probability under negative detuning (red-shifted cavities, where $E_M > \hbar\omega_0)$\cite{Ribeiro2022MultimodeFluctuations}. We investigate the early dynamics in detuned microcavities by computing $v_0$ for a variable $E_M$ and fixed cavity lowest-energy mode $\hbar\omega_0 = 2.0~\text{eV}$. To gain insight into the long-time properties of the wave packet, we also show the maximum RMSD value observed over 5 ps.

\par
\begin{figure*}
    \centering
    \includegraphics[width=\textwidth]{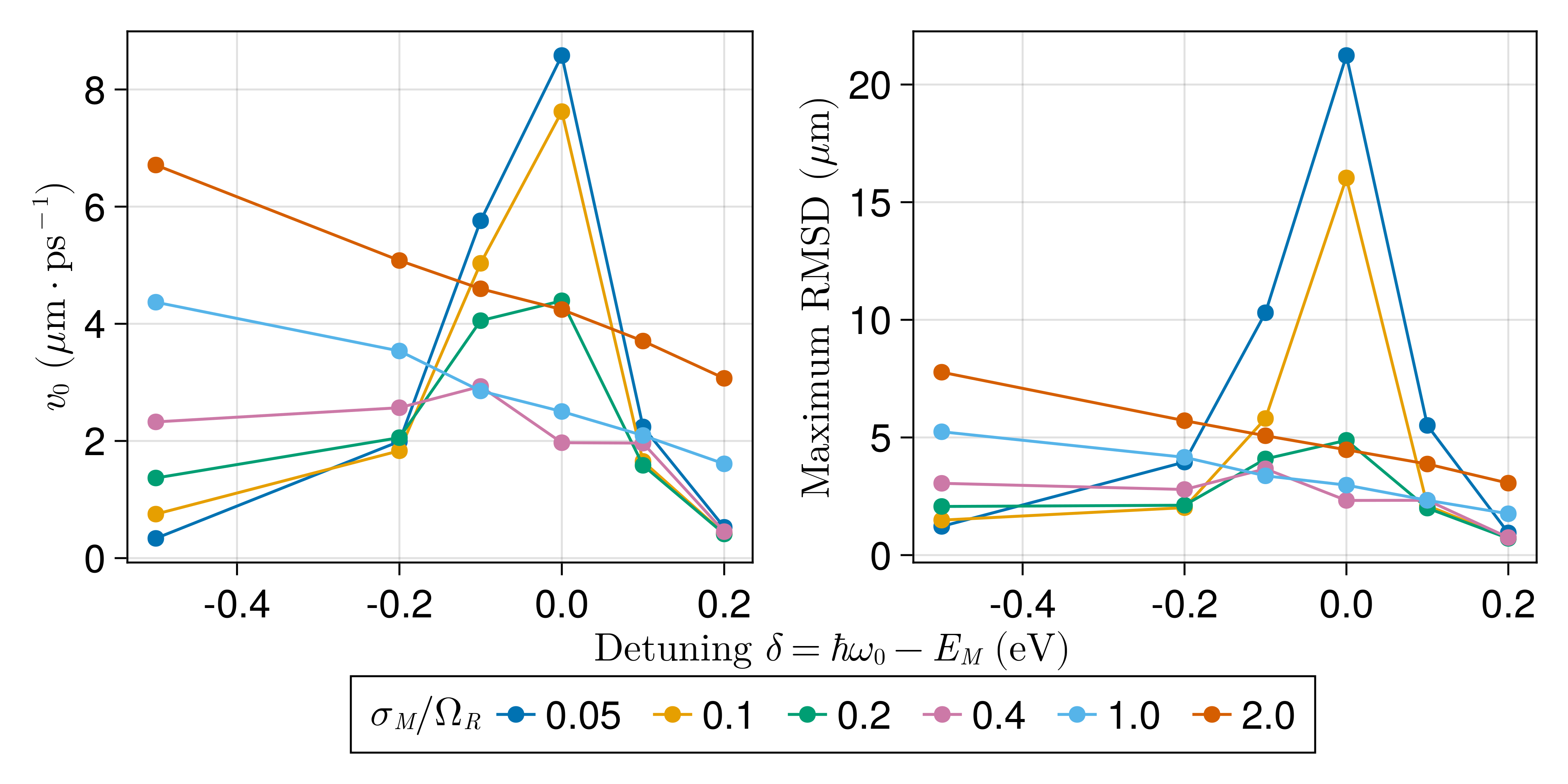}
    \caption{Disorder-dependent detuning effects on coherent exciton transport measured by \textbf{(a)} the initial spread velocity ($v_0$), and \textbf{(b)} the maximum RMSD over 5 ps. In all cases, $\Omega_R = 0.1$ eV and $\sigma_x = 240$ nm.}
    \label{fig:detuning}
\end{figure*}

\begin{figure*}
    \centering
    \includegraphics[width=0.9\textwidth]{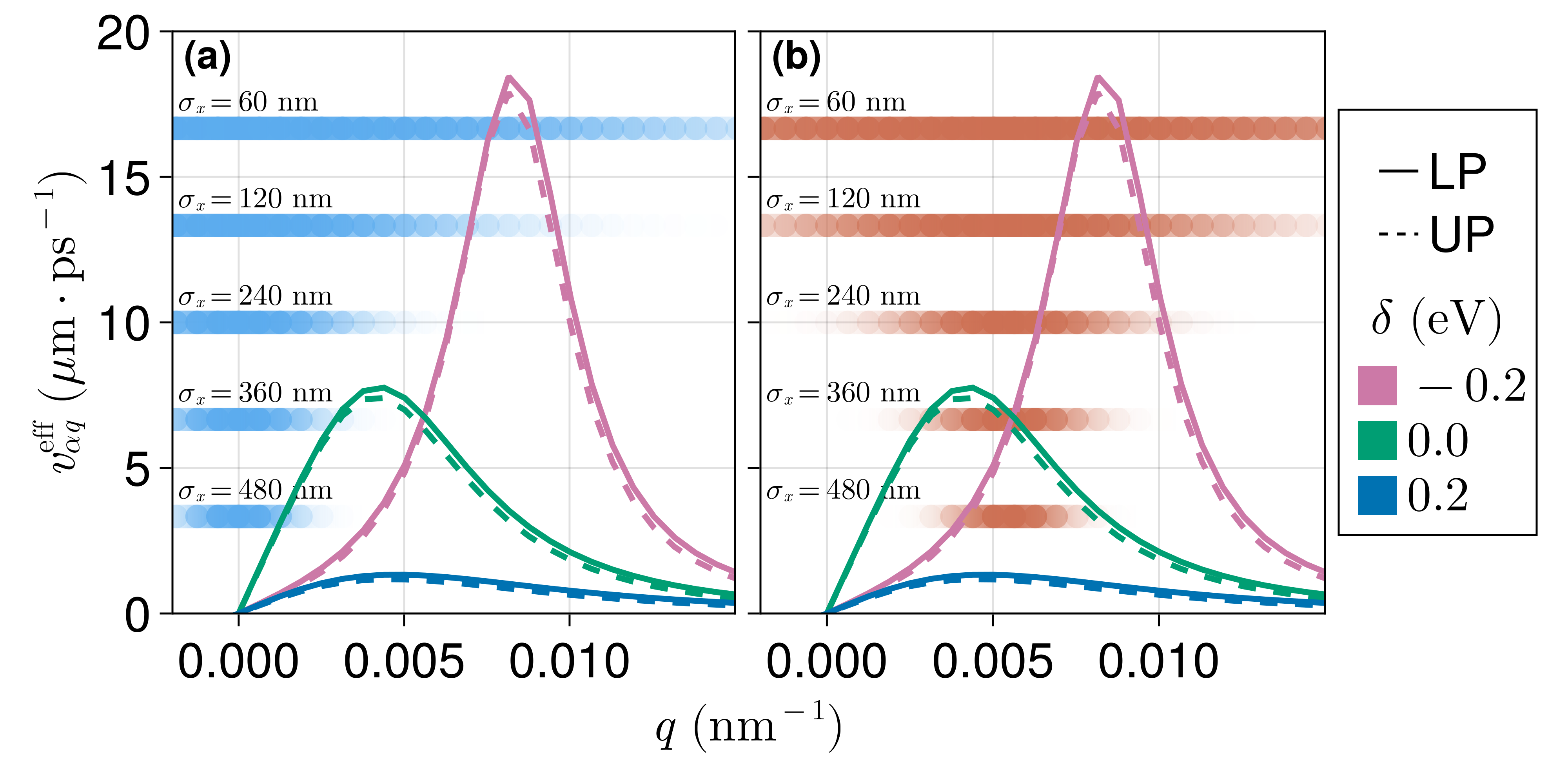}
    \caption{Effective group velocity (Equation \eqref{vgexc}) for variable detuning $\delta = \hbar\omega_0 - E_M$. Horizontal gradient bars represent the $P(q)$ distribution for distinct exciton wave packets (see Equation \eqref{gaussianwvp}) with \textbf{(a)} $\bar{q}_0 = 0.0$ and \textbf{(b)} $\bar{q}_0 = 0.0055$ nm$^{-1}$. The overlap between the gradient bars ($P(q)$) and the $v_{\alpha q}^\text{eff}$ curves yields the exciton spread velocity $v_0$ (Equation \eqref{v02}). See text for more details.}
    \label{fig:vgdetun}
\end{figure*}

 In Figure \ref{fig:detuning}, we find detuning effects on $v_0$ and the maximum RMSD are very similar. Under weak disorder ($\sigma_M/\Omega_R < 0.4$), both $v_0(\delta)$ and RMSD$(\delta)$ are peaked at $\delta = 0$, i.e., the coherent exciton motion is faster when the cavity is in resonance with the dipolar excitation. This can be rationalized with Figure \ref{fig:vgdetun}a which shows a strong dependence of $v_{\alpha q}^{\text{eff}}$ on detuning. Blue-shifted microcavities lead to the slowest exciton motion as evidenced by the consistently smaller $v_{\alpha q}^\text{eff}$ obtained for $\delta = 0.2$ eV in Figure \ref{fig:vgdetun}. On the other hand, red-shifted microcavities have small $v_{\alpha q}^{\text{eff}}$  at $q$ close to zero but higher values (compared to the resonant cavity) at sufficiently large $q$. Since the initial wave packets of Figure \ref{fig:detuning} have $\bar{q}_0 = 0$, the dominant polariton contributions to the evolution are those for which $v_{\alpha q}^{\text{eff}}$ is larger at zero-detuning in comparison to the red-shifted case. Nevertheless, comparison between $v_{\alpha q}^{\text{eff}}$ at zero and negative detuning in Figure \ref{fig:vgdetun}b suggests that polariton-mediated exciton wave packet transport can be much faster in red-shifted cavities when the initial-state is prepared with $\bar{q}_0$ close to the maximum of $v_{\alpha q}^{\text{eff}}$. Indeed, we show numerical results in our supporting information (Figure S13) that confirm this prediction.

\par In the presence of stronger disorder ($\sigma_M > 0.4 \Omega_R)$, Figure \ref{fig:detuning} shows the condition $\delta=0$ no longer provides a maximum RMSD and $v_0$, and the optimal detuning value is shifted toward negative values. Therefore, under sufficient disorder, red-shifting the microcavity enhances the exciton ballistic transport and the transport distance regardless of the initial wave packet preparation. This feature may be understood by noting that in the presence of a significant amount of static disorder, many dipoles will have excitation energies below the lowest-energy microcavity mode. In this case, it becomes advantageous to employ negative detuning, since it supports lower energy photon modes that can interact resonantly with the dipoles with lower excitation energy. Conversely, raising $\hbar\omega_0 - E_M$ to positive values leads to a reduction in the light-matter spectral overlap and, therefore, the maximum RMSD and $v_0$ consistently decrease as the microcavity is blue-shifted ($\delta > 0$). Despite an optimum not being visible in Figure 7, we show in Figure S14 that both $v_0$ and the maximum RMSD reach a plateau between $\delta = -1$ eV and $\delta = -2$ eV. However, we highlight that under such extreme values of negative detuning, the energy gap between the dipole excitation ($E_M$) and higher photonic bands (e.g., $n_y = 1$, $n_z = 2$ and vice-versa) is about the same as the gap between the dipole excitation and the lowest photon mode ($q =0$,  $n_y = n_z = 1$). In this case, higher energy photon bands may start contributing to the dynamics just as much as EM modes in the $n_y = n_z = 1$ band. Therefore, a systematic analysis at large negative detunings must include multiple microcavity bands. We present these results here as they show a well-defined theoretical limit for the employed single-band model.

\subsection*{Losses and dynamical disorder}
We finalize our discussion by offering brief considerations of how the exciton transport phenomena examined in this article would be affected by photon leakage and dynamical disorder (e.g., represented by time-dependent stochastic fluctuations of excitonic transition energies, dipole orientations or intersite distances).

Photon leakage through imperfect mirrors is an unavoidable characteristic of physics in optical microcavities\cite{kavokin2017microcavities}. However, the impact of photon losses on the polariton-assisted exciton transport discussed here is minimal, especially at moderate and large disorder. Under these conditions, the typical time-dependent photon content is small enough (see Figures 2 and S1) that wave packet decay by photon loss is a slow process. For example, if the empty cavity photon lifetime is 50 fs, the corresponding wave packet lifetime is at least one or two orders of magnitude bigger, depending on its photonic content. Therefore, while cavity leakage could prevent observation of DET and subsequent Anderson localization in moderate and strongly disordered systems, the ultrafast ballistic regime would be almost unaffected.

In the weakly disordered scenario where photon content fluctuations can reach a significant value (Figures 2 and S1), the fast decay of the wave packet would prevent efficient energy transport. Nevertheless, even in this case, incorporation of cavity losses is unlikely to change the main qualitative trends observed for $v_0$ as this quantity is computed at ultrafast times (e.g., Figures S10-S12 show that the behavior of $v_0$ with varying disorder and Rabi splitting are independent of the linear fit cutoff time).

The interplay between static and dynamical disorder is an interesting topic that we aim to address in detail in future work. Here, we simply point out some expected effects of dynamical disorder on the examined polariton-assisted exciton transport. First, dynamical disorder suppresses Rabi oscillations and may slow down the relatively efficient ultrafast exciton transport we observe at weak disorder. On the other hand, recent work by Cui and Nitzan has shown that dynamical disorder boosts the early-time exciton transport\cite{cui2023}. Further, it is well known that dynamical disorder prevents Anderson localization and leads to diffusive behavior at long times\cite{Moix2013}. The latter behavior is generic so we expect it to persist in the moderate and strong (static) disorder regimes of polariton-assisted exciton transport. A recent study\cite{Sokolovskii2023} suggests that excitonic interaction with a dynamical (finite-temperature) environment shortens the ballistic transport regime time interval and leads to a quick crossover to diffusive behavior. We expect similar features will emerge when dynamical disorder is incorporated into our model, but future work is necessary to determine the interplay between the effects of static and dynamical disorder in the investigated polariton-assisted exciton transport.

\section*{Conclusions}
\par We examined coherent polariton-mediated exciton transport on a lossless disordered polaritonic wire. Our analysis shows that the initial exciton wave packet (i.e., its spread and average momentum) strongly influences its ballistic propagation regime and may be optimized to maximize its early mobility. A striking contrast between polariton-mediated and purely excitonic transport was also noted here. Previous work showed that short-time direct exciton energy transport (via dipole-dipole interactions) is enhanced by disorder \cite{cui2023}. Here, we find, contrarily, that disorder systematically suppresses the initial wave packet spread. This implies a fundamental distinction in how disorder impacts coherent exciton energy transport inside and outside an optical microcavity.

\par We also analyzed the interplay of detuning and static disorder as factors impacting the ballistic transport regime. We found that while blue-shifted cavities always presented a slower exciton wave packet transport, red-shifted microcavities showed richer behavior, i.e., both suppression and enhancement of transport can be attained depending on the level of disorder and the initial state preparation. 

To rationalize these results, we introduced the effective exciton group velocity $v_{\alpha q}^\text{eff}$, which can be computed from the system dispersion and the excitonic content of each polariton eigenmode. The early ballistic transport can be estimated by combining $v_{\alpha q}^\text{eff}$ with the initial exciton probability distribution in $q$ space. This analysis leads to a design principle for optimizing the initial exciton state for enhanced ultrafast coherent transport based on the complex interplay between disorder and tunable light-matter parameters such as detuning, Rabi splitting, and initial wave packet width and momentum. The optimal initial state for exciton transport must have: i) an initial wave vector matching the maximum value of the effective exciton group velocity and ii) a sufficiently narrow spread in $q$-space such that it only spans eigenmodes with large effective exciton group velocity.

\par Our theoretical analysis of exciton wave packet propagation in terms of the newly introduced effective exciton group velocity led to qualitative agreement with simulations even under moderate disorder ($\sigma_M/\Omega_R = 0.4$). Such robustness and generalizability suggest our results will be useful for future theoretical and experimental studies of transport in optical cavities.

\begin{acknowledgement}
R.F.R. acknowledges generous start-up funds from the Emory University Department of Chemistry. 
\end{acknowledgement}

\bibliography{refs}

\clearpage
\begin{center}
    \centering
    \LARGE \textbf{Supporting Information} \\Coherent transient exciton transport in disordered polaritonic wires \\
    \Large
    Gustavo J. R. Aroeira, Kyle T. Kairys, and Raphael F. Ribeiro$^*$ \\[3mm]
    \normalsize
    \textit{Department of Chemistry and Cherry Emerson Center for Scientific Computation, Emory University, Atlanta, Georgia 30322, United States of America} \\
    \ttfamily \small Email: raphael.ribeiro@emory.edu    
\end{center}
\addcontentsline{toc}{section}{Supporting Information}

\renewcommand{\thefigure}{S\arabic{figure}}
\renewcommand{\thetable}{S\arabic{table}}
\begin{table}[h]
    \centering
    \begin{tabular}{c l p{20mm}}
     \textbf{Symbol} & \textbf{Description} & \textbf{Value}   \\[1mm]
     \hline
     $N_M$  & Number of dipoles in the wire. & 5000 \\
     $N_c$  & Number of cavity modes used to describe the radiation field inside the cavity. & 1001 \\
     $\Omega_\text{R}$ & Rabi splitting: a measure of the collective light-matter interaction strength. & Variable \\
     $a$ & Intersite distance. & 10 nm \\
     $\sigma_\text{a}$ & Standard deviation of sites positions. & 1 nm \\
     $E_\text{M}$ & Dipole excitation energy. & Variable \\
     $\sigma_\text{M}$ & Standard deviation of the distribution of dipole excitation energies. & Variable \\
     $L_z$ & Wire length along $z$ dimension. & 0.4 $\mu$m \\
     $L_y$ & Wire length along $y$ dimension. & 0.2 $\mu$m \\
     $L_x$ & Wire length along $x$ dimension. & 50 $\mu$m \\
     $\epsilon$  & Relative static permittivity. & 3 \\
     $n_y$ & Cavity quantum number associated with the $y$ dimension. & 1 \\
     $n_z$ & Cavity quantum number associated with the $z$ dimension. & 1 \\
     $m_x$ & Cavity quantum number associated with the $x$ dimension. & $\in \mathbb{Z}$ \\
     $q$ & $x$-component of the wavevector $\mathbf{k}$ & $2\pi m_x/L_x$. \\
     $\sigma_x$ & Initial wave packet width. & Variable \\
     $\bar{q}_0$ & Average initial exciton momentum along $x$ & Variable \\[1mm]
     \hline
    \end{tabular}
    \caption{List of symbols and notation used in this work}
    \label{tab:symbols}
\end{table}

\clearpage

\setcounter{figure}{0}    

\newpage

\section{RMSD for a traveling exciton-polariton}

\subsection{Preliminary expressions}

\subsubsection{Wave packet representations}
Before we proceed to our main problem, let us establish some intermediate results that will be useful later. First, consider the wave packet at $t = 0$ represented in the position basis of $N_M$ dipoles
\begin{align}
    \ket{\psi_0} = \sum_{n=1}^{N_M} c_n \ket{n} \;.
\end{align}
Therefore, the probability of finding the $n$-th dipole excited is $|c_n|^2$. Note that in the main text, we use the notation $\ket{n;0}$ to represent the state where the $n$-th dipole is excited and no photon modes are populated. For simplicity, we use $\ket{n;0}\rightarrow\ket{n}$ here as the photonic degrees of freedom will not appear explicitly in the following expressions. We can transform $\ket{\psi_0}$ into its $k$-space representation using the resolution of the identity (in the exciton Hilbert space) $\sum_k \ketbra{k}{k} = \id_E$. That is,
\begin{align}
    \ket{\psi_0} &= \sum_k \sum_{n=1}^{N_M} c_n\braket{k}{n}\ket{k} \;, \\
    &= \sum_k \prts{\frac{1}{\sqrt{N_M}}\sum_{n=1}^{N_M} c_n e^{-ikx_n} }\ket{k} \;,
\end{align}
where we have used $\braket{k}{n} = \frac{e^{-ikx_n}}{\sqrt{N_M}}$. Defining
\begin{align}
    c_k = \frac{1}{\sqrt{N_M}}\sum_{n=1}^{N_M} c_n e^{-ikx_n} \label{ck}
\end{align}
we have the wave function in $k$-representation
\begin{align}
    \ket{\psi_0} = \sum_k c_k \ket{k} \;.
\end{align}
From the expression above, we see that the initial exciton probability distribution in wave number space is $P_k = |c_k|^2$. 

\subsubsection{Polariton Eigenstates}

In the case where the dipole ensemble is translationally invariant, we can assign each eigenmode to the upper polariton (UP) branch if its energy is above the dipolar transition energy or lower polariton (LP) otherwise. The general Hamiltonian for the system can be block diagonalized using the transformation in Eq. \ref{ck}. The resulting Hamiltonian is a direct sum of $2\times 2$ Hamiltonians where the $k$ wave number is preserved, that is, there is no mixing of different values of $k$. Using $q$ for the wave number of the polariton states we can write
\begin{align}
    \hat{H} \ket{\alpha q} = \hbar\omega_{\alpha q} \ket{\alpha q} \;, \label{H}
\end{align}
where $\alpha$ is LP or UP. Moreover, the amplitude of the $n$-th dipole on the eigenstate $\ket{\alpha q}$ is
\begin{align}
    \braket{n}{\aq} = -i\sqrt{\frac{\piaq}{N_M}}e^{iqx_n} \;, \label{mol_cont}
\end{align}
where $\piaq = \sum_{n=1}^{N_M} \braket{\aq}{n}\braket{n}{\aq}$ is the total exciton content of the eigenstate $\ket{\aq}$. With this result, the overlap of each eigenstate with the initial wave packet is given by
\begin{align}
    \braket{\psi_0}{\aq} &= \sum_{n=1}^{N_M} c^*_n\braket{n}{\aq} \\
    &= -i\sqrt{\piaq} \prts{ \frac{1}{\sqrt{N_M}} \sum_{n=1}^{N_M} c^*_n e^{iqx_n} } \;.
\end{align}
The term in brackets above is the complex conjugate of Eq. \eqref{ck}. Thus,
\begin{align}
    \braket{\psi_0}{\aq} = -i \sqrt{\piaq} c^*_{q} \;. \label{psi0aq}
\end{align}
The resolution of the identity in the eigenmode basis is given by
\begin{align}
    \id = \sum_\alpha \sum_q \ket{\aq}\bra{\aq}\;. \label{reseigen}
\end{align}

\subsubsection{Continuum limit}
In what follows, we will take the thermodynamic limit where $L$ and $N_M$ go to infinity at the same rate, so $L/N_M$ is fixed. This will lead to closed-form expressions for excitonic observables in the absence of disorder. We will assume throughout that all relevant functions (of the dipole position $x_n = na$ and wavenumber $q$) are sufficiently slowly varying at the scale of the spatial lattice constant $x_{n+1}- x_n = \Delta x = a$ and reciprocal lattice spacing $\Delta q  = 2\pi/L$ so the continuum limit is well defined. We also assume the wave packet vanishes sufficiently fast outside a finite closed subset of position or wave number space. Under these conditions, any sum over discrete functions of the dipole position $x_n$ can be replaced by an integral over all space following
\begin{align}
    \sum_{n=1}^{N_M} f_n \rightarrow \frac{1}{a}\int f(x)\mathrm{d}x,
\end{align}
where $f(x)$ is the continuum representation of $f_n$ satisfying $f(x_n) = f_n$. Likewise, the continuum limit for sums of discrete functions of $q$ is obtained from
\begin{align}
   \sum_q g_q \rightarrow  \frac{L}{2\pi}\int~g(q)\mathrm{d}q,
\end{align}
where $g(q)$ is the continuum representation of $g_q$. 

The continuum limit of the resolution of the identity in the eigenmode basis is given by
\begin{align}
     \id = \frac{L}{2\pi} \sum_\alpha \int \ket{\aq}\bra{\aq}dq\;
\end{align}
Thus, it is convenient to redefine $\ket{\aq}$ following
\begin{align}
    \ket{\aq} \rightarrow \sqrt{\frac{2\pi}{L}}\ket{aq}, \label{alphaq_con} 
\end{align}
so as to recover the identity operator in the standard form
\begin{align}
       \id =  \sum_\alpha \int \ket{\aq}\bra{\aq}dq\; \label{contID}
\end{align}
Similar manipulations can be performed for the states living in the matter Hilbert space, e.g.,
\begin{align}
   & \id_E  = \sum_{n=1}^{N_M} \ket{n}\bra{n} \rightarrow \frac{1}{a}\int\ket{x}\bra{x} \mathrm{d}x, \\
   &\id_E = \sum_{q} \ket{q}\bra{q} \rightarrow \frac{L}{2\pi}\int \ket{q}\bra{q}\mathrm{d}q.
\end{align}
These identities suggest redefining the matter states in the continuum limit following
\begin{align}
&\ket{n} \rightarrow \sqrt{a}\ket{x}, \\
&\ket{q} \rightarrow \sqrt{\frac{2\pi}{L}} \ket{q}.
\end{align}
Using these relations, it follows in the continuum limit
\begin{align}
    \braket{k}{x} = \sqrt{\frac{L}{2\pi a}}\braket{k}{n} = \frac{e^{-ik x}}{\sqrt{2\pi}},
\end{align}
From these results, we obtain the wave packet representation in position and wave number space analogous to those of Sec. 1.1.1.
\begin{align}
& \ket{\psi_0} = \int c(x) \ket{x} \mathrm{d}x, ~~ c(x) = \braket{x}{\psi_0} \\
& \ket{\psi_0} = \int \int c(x) \braket{q}
{x}~ \ket{k} \mathrm{d}k\mathrm{d}x = \int c(k) \ket{k}\mathrm{d}k,
\end{align}
where the wave vector amplitude in the continuous wave number space is given by
\begin{align}
    c(k) = \frac{1}{\sqrt{2\pi}}\int c(x) e^{-ikx}\mathrm{d}x.
\end{align}
By performing the analogous transformations to the polariton eigenmodes we obtain
\begin{align}
& \braket{x}{\aq} = -i\sqrt{\frac{\Pi_{\aq}}{2\pi}} e^{i q x}, \\
& \braket{\psi_0}{\aq} = -i \sqrt{\Pi_{\aq}} c^*(q).
\end{align}
\subsubsection{Time-evolved wave packet}

The wave packet at an arbitrary time $\ket{\psi_t}$ can be written in terms of the initial state using the time-evolution operator
\begin{align}
    \ket{\psi_t} = e^{-i\hat{H}t/\hbar} \ket{\psi_0} \;.
\end{align}
Using the eigenstates discussed in the previous section, in particular Eq. \ref{contID}, we can resolve the time evolution as
\begin{align}
    \ket{\psi_t} = \sum_\alpha \fullint e^{-i\hat{H}t/\hbar} \ket{\aq}\braket{\aq}{\psi_0} dq \;.
\end{align}
Using Eq. \ref{H} and \ref{psi0aq} we get
\begin{align}
    \ket{\psi_t} = i \sum_\alpha \fullint \sqrt{\piaq} c(q) e^{-i\omega_\aq t} \ket{\aq} dq \;. \label{timeevol}
\end{align}

\subsection{Exciton mean squared displacement}
The exciton mean squared displacement $\Delta x_E^2$ can be written as
\begin{align}
    \Delta x_E^2(t) &= \frac{1}{P_M(t)} \int|\braket{x}{\psi_t}|^2 (x-x_0)^2 \mathrm{d}x \\
    P_M(t) &= \int|\braket{x}{\psi_t}|^2 \mathrm{d}x  \;,
\end{align}
where $x_0$ is the center of the initial wave packet. We start by computing $P_M(t)$ from the inner product of $\ket{x}$ with $\ket{\psi_t}$ (Eq. \ref{timeevol}). The result is
\begin{align}
    \braket{x}{\psi_t} = \frac{1}{\sqrt{2\pi}}\sum_\alpha \fullint \piaq c(q) e^{-i\omega_\aq t} e^{iqx} dq \;. \label{npsit}
\end{align}
Hence,
\begin{align}
    P_M &= \int  \braket{\psi_t}{x}\braket{x}{\psi_t}\mathrm{d}x \\
    &= \frac{1}{2\pi}\sum_\alpha \sum_\beta \fullint \fullint c^*(q')c(q)\pibqp\piaq e^{-i(\omega_\aq-\omega_\bqp) t} \left(\int e^{i(q-q')x} dx\right) dq dq' \;.
\end{align}
Using $2\pi\delta(q-q') = \int~\text{exp}[i(q-q')x]dx$, we obtain
\begin{align}
    P_M &= \sum_\alpha \sum_\beta \fullint |c(q)|^2 \pibq\piaq e^{-i(\omega_\aq-\omega_\bq) t}  dq  \;.
\end{align}
The double sum over $\alpha$ and $\beta$ produces four terms. Those are
\begin{align}
    P_M &= \fullint |c(q)|^2 \prts{\Pi^2_{\text{LP}q} + \Pi^2_{\text{UP}q} + \Pi_{\text{LP}q}\Pi_{\text{UP}q}e^{-i(\omega_{\text{LP}q}-\omega_{\text{UP}q}) t} + \Pi_{\text{LP}q}\Pi_{\text{UP}q}e^{-i(\omega_{\text{UP}q}-\omega_{\text{LP}q}) t}}  dq \\
    &= \fullint |c(q)|^2 \prts{\Pi^2_{\text{LP}q} + \Pi^2_{\text{UP}q} + 2\Pi_{\text{LP}q}\Pi_{\t{UP}q}\cos{[(\omega_{\text{UP}q}-\omega_{\text{LP}q}) t]}}  dq \;.
\end{align}
As described in the main text, we measure the exciton spread velocity ($v_0$) by using a linear fit over the initial 500 fs. This process averages out the oscillating terms observed above. From now on, we ignore the time-dependent fluctuations of $P_M(t)$ and work with the more relevant time-averaged exciton content
\begin{align}
    \bar{P}_M = \fullint |c(q)|^2 \prts{\Pi^2_{\t{LP}q} + \Pi^2_{\t{UP}q}} dq \label{SI_PM} \;.
\end{align}

\par Next, we analyze the remaining part of $\Delta x_E^2$. For the sake of simplicity and without loss of generality we assume $x_0 = 0$. We can use the results from Eq. \ref{npsit} to write
\begin{align}
    P_M \Delta x_E^2(t) &= \int |\braket{x}{\psi_t}|^2 x^2 dx\\
    &= \frac{1}{2\pi} \sum_\alpha \sum_\beta \fullint \fullint c^*(q')c(q)\pibqp\piaq e^{-i(\omega_\aq-\omega_\bqp) t} \prts{\int x^2 e^{i(q-q')x}dx} dq dq' \;.
\end{align}
Rearranging the integrals and using the substitution $x^2e^{i(q-q')x} = -\dqq~ {e^{i(q-q')x}}$, we obtain
\begin{align}
    P_M  \Delta x_E^2(t) &= -\frac{1}{2\pi} \sum_\alpha \sum_\beta \fullint c(q)\piaq e^{-i\omega_\aq t} \prts{\fullint c^*(q')\pibqp e^{i\omega_\bqp t} \dqq{}\prts{\int e^{i(q-q')x} dx} dq'} dq \\
    & = -\sum_\alpha \sum_\beta \fullint c(q)\piaq e^{-i\omega_\aq t} \prts{\fullint c^*(q')\pibqp e^{i\omega_\bqp t} \dqq{} \delta(q-q')dq'} dq\; \label{a}
\end{align}
Using the smoothness and compact support assumptions specified in our discussion of the continuum limit, it follows that
\begin{align}
    \fullint f(q') \dqq{} \delta(q-q') dq' = \fullint  \dqq{f(q')} \delta(q-q') dq' = \dqq{f(q)} \;.
\end{align}
Therefore, we can reduce Eq. \ref{a} to the form
\begin{align}
    P_M \Delta x_E^2(t) &= - \sum_\alpha \sum_\beta \fullint c(q)\piaq e^{-i\omega_\aq t} \dqq{}\left[c^*(q)\pibq e^{i\omega_\bq t}\right] dq \;. \label{b}
\end{align}
To proceed, we evaluate the second derivative with respect to $q$ 
\begin{align}
    e^{-i\omega_{\beta q}t}\dqq{}\left[c^*(q)\pibq e^{i\omega_\bq t}\right] & = e^{-i\omega_{\beta q}t}\pder{}{q}\left[\pder{c^*(q)\pibq}{q}e^{i\omega_\bq t}+ {-i\pder{\omega_{\bq}}{q} t} c^*(q)\pibq e^{i\omega_\bq t}\right] \nonumber \\
    & = \dqq~{[c^*(q)\pibq]} -2it\pder{\omega_{\bq}}{q}\pder{}{q}[c^*(q)\pibq]  -tc^*(q)\pibq\left[i\dqq{\omega_{\bq}}   +  t\left(\pder{\omega_{\bq}}{q}\right)^2 \right] \label{c}.
\end{align}
To obtain a more compact expression, we define
\begin{align}
    \Delta_{\beta \alpha}(q) &= \omega_\bq - \omega_\aq, \\
    \gamma_{\aq} &= c(q) \piaq, \\
    \gamma_{\bq}^* &= c^*(q) \pibq, \\
    v_g^{\bq} &= \pder{\omega_\bq}{q}.
\end{align}
Using these definitions and plugging Eq. \ref{c} into Eq. \ref{b} we get 
\begin{align}
    P_M \Delta x_E^2(t) &= - \sum_\alpha \sum_\beta \fullint \prts{\dqq{\gamma^*_{\bq}} -it \pder{v_g^{\bq}}{q}\gamma_{\bq}^* -2iv_g^{\bq}t \pder{~\gamma^*_{\bq}}{q} - (v_g^{\bq}t)^2 \gamma^*_{\bq}}\gamma_{\aq}e^{i\Delta_{\beta\alpha}(q)t} dq 
\end{align}
Since our goal is to obtain the term of $\sqrt{\Delta x_E^2}$ proportional to $t$, we will ignore the cross LP-UP oscillating terms where $\beta \neq \alpha$ (note this is consistent with our neglect of oscillations in the previous treatment of $P_M$). It follows in this case that
\begin{align}
    P_M \Delta x_E^2(t) &= - \sum_\alpha  \fullint \left[\left(\dqq{~}\gamma^*_{\aq}\right)\gamma_{\aq} -it \left(\pder{~}{q}v_g^{\aq}\right)|\gamma_{\aq}|^2 -2itv_g^{\aq} \left(\pder{~}{q}\gamma^*_{\aq}\right)\gamma_{\aq} - (v_g^{\aq}t)^2 |\gamma_{\aq}|^2\right]  dq \; \label{d}
\end{align}
Note from the above expression that the time-independent contribution gives the exciton spread at $t = 0$ i.e.,
\begin{align}
    P_M \Delta x_E^2(0) = -\sum_\alpha  \fullint \left(\dqq{~}\gamma^*_{\aq}\right)\gamma_{\aq}dq.
\end{align}
It can be shown with integration by parts that the $\mathcal{O}(t)$ term in Eq. \ref{d} is given by
\begin{align}
itv_{g}^{\aq}\left(\frac{\partial}{\partial q}\gamma_{\aq}\right)\gamma_{\aq}^* - itv_{g}^{\aq}\left(\frac{\partial}{\partial q}\gamma_{\aq}^{*}\right)\gamma_{\aq} = -2v_g^{\aq}t\times\text{Im}\left[\gamma_{\aq}^*\frac{\partial}{\partial q} \gamma_{\aq}\right]  \label{otterm}
\end{align}
Hence, it can be seen that if $\gamma_{\aq} = c_q \piaq$ is real this term vanishes. In the case of Gaussian wave packets in the continuum limit, we have
\begin{align}
    c(q) = \frac{1}{\sqrt{2\pi}}\fullint c(x) e^{-iqx} dx = \frac{1}{2\pi\sigma_x}\fullint e^{-\frac{(x - x_0)^2}{2\sigma_x}}  e^{i(\bar{q}_0 - q) x} dx \\[2mm]
    \frac{1}{2\pi\sigma_x}\fullint e^{-\frac{(x - x_0)^2}{2\sigma_x}}  [\cos{(\bar{q}_0 - q) x} + i\sin{(\bar{q}_0 - q) x}] dx \;.
\end{align}
Due to parity considerations, the sine part of this expression vanishes and $c(q)$ must be real. Consequently, Eq. \ref{otterm} is zero for Gaussian wave packets.

\par It follows that the ballistic spread velocity $v_0$ for a Gaussian wave packet is given by 
\begin{align}
    v_0^2 = \frac{\Delta x_E^2(t) - \Delta x_E^2(0)}{t^2} = \frac{1}{P_M}\fullint |c(q)|^2 \left[ \prts{v_g^{\t{LP}}(q) \Pi_{\t{LP}q}}^2 + \prts{v_g^{\t{UP}}(q) \Pi_{\t{UP}q}}^2 \right] dq \;.
\end{align}
Defining an effective group velocity as $v_\aq^\t{eff} = v_g^{\alpha}(q) \piaq$ and disregarding the (here constant) prefactor $P_M$ we arrive at the following final estimate for the observed ballistic exciton spread velocity
\begin{align}
    v_0^2 \approx \fullint |c(q)|^2 \left[ (v_{\t{LP}q}^\t{eff})^2 + (v_{\t{UP}q}^\t{eff})^2 \right] dq \;.
\end{align}
\clearpage

\section{Exciton migration probability}
In this section, we obtain insight into the exciton migration probability defined in the main text as
\begin{align}
\chi(t) = 1 - \frac{1}{P_M(t)}\sum_{n\in \mathcal{I}} |\braket{n}{\psi(t)}|^2, \label{eq:chit}
\end{align}
where $\mathcal{I}$ is the integer interval $[n_{\text{min}},n_{\text{max}}]$ containing the indices of the dipoles comprising $99\%$ of the initial  wave packet probability. For the sake of simplicity, we ignore $P_M(t)$ below since both at very early times $t\rightarrow 0$ and late times $t\rightarrow \infty$, $P_M(t)$ is approximately constant. Let $P_n(t)$ denote the time-dependent probability to to detect an exciton at the $n$th site and $\{\ket{A},\ket{B},...\}$ correspond to the set of eigenstates of $H$ with eigenvalues $E_A = \hbar \omega_A$, etc. It follows that we can write the probability $P_n(t)$ to detect an exciton at the $n$th site at time $t$ is given by
\begin{align}
    P_n(t) & = |\braket{n}{\psi(t)}|^2 = \left|\sum_{A}\braket{n}{A}\braket{A}{\psi_0} e^{-i\omega_A t}\right|^2 \nonumber \\
    & = \left|\sum_{A} \sum_{m_1} A_n A_{m_1}^* c_{m_1} e^{-i\omega_A t}\right|^2\nonumber\\
    & = \sum_{A,B} \sum_{m_1, m_2} A_n A_{m_1}^* B_n^* B_{m_2} c_{m_1} c_{m_2}^* e^{-i(\omega_A-\omega_B) t}.
\end{align}
where $A_n$ corresponds to the probability amplitude to detect an exciton at site $\ket{n}$ when the system is in eigenstate $\ket{A}$, and $c_n = \braket{n}{\psi_0}$ as in the previous section. We can decompose $P_n(t)$ into 
\begin{align}
    P_n(t) = P_n + \Delta P_n (t),
\end{align}
where $P_n$ corresponds to the time-independent asymptotic part and $\Delta P_n(t)$ is the time-fluctuating contribution to $P_n(t)$. Each of these terms is given explicitly by
\begin{align}
& P_n = \sum_{A} \sum_{m_1,m_2} |A_n|^2 A_{m_1}^*A_{m_2} c_{m_1} c_{m_2}^*, \\
& \Delta P_n(t) = \sum_{A}\sum_{B\neq A}\sum_{m_1,m_2} A_n A_{m_1}^* B_{n}^* B_{m_2} c_{m_1} c_{m_2}^* e^{-i(\omega_A-\omega_B)t}.
\end{align}
Given the definition of $\chi(t)$ in Eq. \ref{eq:chit}, it follows that we may also define an asymptotic time-independent part contribution $\chi$ and an oscillatory contribution $\Delta \chi(t)$ such that $\chi(t) = \chi + \Delta \chi(t)$ given by
\begin{align}
    & \chi \approx 1-\sum_{A} \sum_{n \in \mathcal{I}} \sum_{m_1,m_2} |A_n|^2 A_{m_1}^* A_{m_2} c_{m_1} c_{m_2}^*, \\
    & \Delta \chi(t) \approx - \sum_{A}\sum_{B\neq A}\sum_{n \in \mathcal{I}}\sum_{m_1,m_2} A_n A_{m_1}^* B_{n}^* B_{m_2} c_{m_1} c_{m_2}^* e^{-i(\omega_A-\omega_B)t},
\end{align}
where the approximate character of the identities emphasizes our neglect of $P_M(t)$ in the definition of $\chi(t)$. We examine $\chi$ and $\Delta\chi(t)$ next starting with the time-independent term which we rewrite as
\begin{align}
\chi & = 1-\sum_{A}\sum_{n\in \mathcal{I}} |A_n|^4 |c_n|^2 - \sum_{A}\sum_{n\in\mathcal{I}} \sum_{m\neq n} |A_n|^2 |A_m|^2 |c_m|^2 \nonumber \\
& - \sum_{A} \sum_{n \in \mathcal{I}} \sum_{m\neq n} |A_n|^2\left(A_{n}^*A_m c_n c_{m}^*+A_{m}^*A_n c_m c_{n}^*\right) 
- \sum_{A}\sum_{n\in\mathcal{I}}\sum_{m_1\neq n}\sum_{m_2\neq m_1,n} |A_n|^2 A_{m_1}^* A_{m_2} c_{m_1} c_{m_2}^*.
\end{align}
The fluctuating term can be written likewise as
\begin{align}
    \Delta \chi(t) = & -\sum_{A}\sum_{B\neq A} \sum_{n\in \mathcal{I}} |A_n|^2 |B_n|^2 |c_n|^2 \t{cos}[(\omega_A-\omega_B)t]  \nonumber \\
   &  - \sum_{A}\sum_{B\neq A}\sum_{n \in \mathcal{I}}\sum_{m \neq n} \l(A_n A_m^* B_n^* B_m |c_m|^2 + |A_n|^2 B_n^* B_m c_n c_m^* +A_n A_m^* |B_{n}|^2 c_m c_n^*\r) e^{-i(\omega_A-\omega_B)t} \nonumber \\
   & - \sum_{A}\sum_{B\neq A}\sum_{n \in \mathcal{I}}\sum_{m_1 \neq n}\sum_{m_2\neq m_1,n} A_n A_{m_1}^* B_{n}^* B_{m_2} c_{m_1} c_{m_2}^* e^{-i(\omega_A-\omega_B)t}.
\end{align}
\par Note that
\begin{align}
    \frac{\mathrm{d}}{\mathrm{d}t}\chi(t) = \frac{\mathrm{d}}{\mathrm{d}t}\Delta\chi(t),
\end{align}
and $\mathrm{d}\Delta\chi(t)/\mathrm{d}t \rightarrow 0$ as $t\rightarrow 0^+$. Therefore, the early growth of $\chi(t)$ is characterized by 
\begin{align}
    G = \lim_{t\rightarrow 0^+}\frac{1}{2}\frac{\mathrm{d}^2}{\mathrm{d}t^2} \Delta \chi(t).
\end{align}

\par \textbf{Strong disorder limit.} To simplify, we perform the disorder  average of $\chi$ which we denote by $\overline{\chi}$ (our notation for disorder average of a quantity $f$ in this section is given by $\overline{f}$). Under sufficiently strong disorder ($\sigma_M \gg \Omega_R$), we expect $\overline{A_{m_1}^* A_{m_2}} \approx |\overline{A_{m_1}}|^2 \delta_{m_1,m_2}$ for any eigenmode $A$, and therefore the disorder-averaged exciton migration probability components satisfy
\begin{align}
    & \overline{\chi}  \approx 1-\sum_{A}\sum_{n\in \mathcal{I}} \overline{|A_n|^4} |c_n|^2 - \sum_{A}\sum_{n\in\mathcal{I}} \sum_{m\neq n} \overline{|A_n|^2 }~\overline{|A_m|^2} |c_m|^2, \label{eq:tichi_d} \\
    & \overline{\Delta \chi(t)} \approx -\sum_{A}\sum_{B\neq A} \sum_{n\in \mathcal{I}}\overline{|A_n|^2|B_n|^2} |c_n|^2 \text{cos}\left[(\omega_A-\omega_B)t\right]. \label{eq:dchi_d}
\end{align}
From the last equation, we directly quantify the  early growth of the migration probability from
\begin{align}
    \overline{G} \approx  \frac{1}{2}\sum_{A} \sum_{B\neq A} \sum_{n\in \mathcal{I}} \overline{|A_n|^2 |B_n|^2} |c_n|^2 (\omega_A-\omega_B)^2. \label{eq:SI_Gd}
\end{align}

\par \textbf{Weak disorder limit.} In the weak disorder case, 
\begin{align}
    G = \frac{1}{2} \sum_{A}\sum_{B\neq A} \sum_{n\in \mathcal{I}} \sum_{m_1,m_2} A_{n}A_{m_1}^* B_{n}^* B_{m_2} c_{m_1} c_{m_2}^* (\omega_A-\omega_B)^2
\end{align}
Assuming only modes with $q \sigma_x \ll 1 $ contribute significantly, we can take the long wavelength limit and ignore the phase difference of eigenmode amplitudes in distinct sites, thus considerably simplifying $G$
\begin{align}
G  \approx \frac{1}{2N_{\mathcal{I}}} \sum_{A}\sum_{B\neq A} \sum_{n\in \mathcal{I}} \sum_{m_1,m_2} |A_n|^2 |B_{n}|^2 (\omega_A-\omega_B)^2, \label{eq:SI_G_nd}
\end{align}
where we made the replacement $c_{m_1}c_{m_2}^* \approx 1/N_{\mathcal{I}}$ and $N_{\mathcal{I}}$ is the number of elements of $\mathcal{I}$. 
\clearpage
\section{Wave packets at long propagation times}

\begin{figure*}
    \centering
    \includegraphics[width=\textwidth]{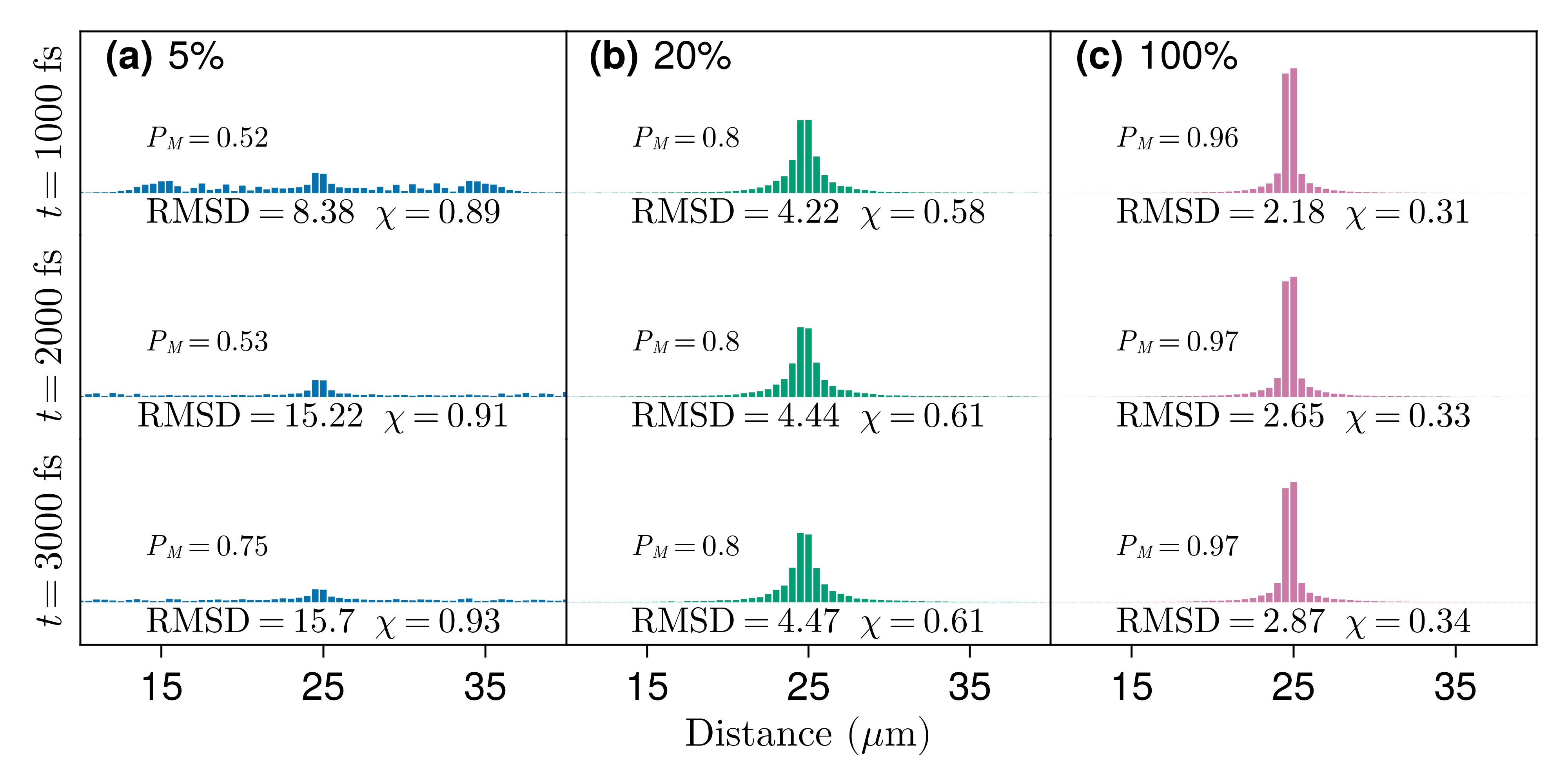}
    \caption{Average exciton wave packet profiles at different time delays and relative disorder strength ($\sigma_M/\Omega_R$) of 5\%, 20\%, and 100\% for (a), (b) and (c), respectively. In all cases, $\Omega_R = 0.1$ eV and the wave packet initial width and mean quasimomentum are $\sigma_x = 120$ nm and $\bar{q}_0 = 0$. Probabilities are grouped in bins containing 50 dipoles spanning 0.5 $\mu m$.  $P_M$, RMSD, and  $\chi$ are defined in Equations \eqref{SI_x2},\eqref{SI_PM}, and \eqref{SI_escape}, respectively. Each profile is obtained from the average of 100 disorder realizations.}
    \label{fig:wvps}
\end{figure*}

Fig. \ref{fig:wvps} shows the shape of wave packets at 1, 2, and 3 ps. We observe that, except when $\sigma_M/\Omega_R = 5\%$, the wave packet remains localized around its initial region. Therefore, the asymptotic behavior seen in Figure 3 (main text) can be attributed to Anderson localization of the excitonic wave packet. 

\begin{figure}
    \centering
    \includegraphics[width=\textwidth]{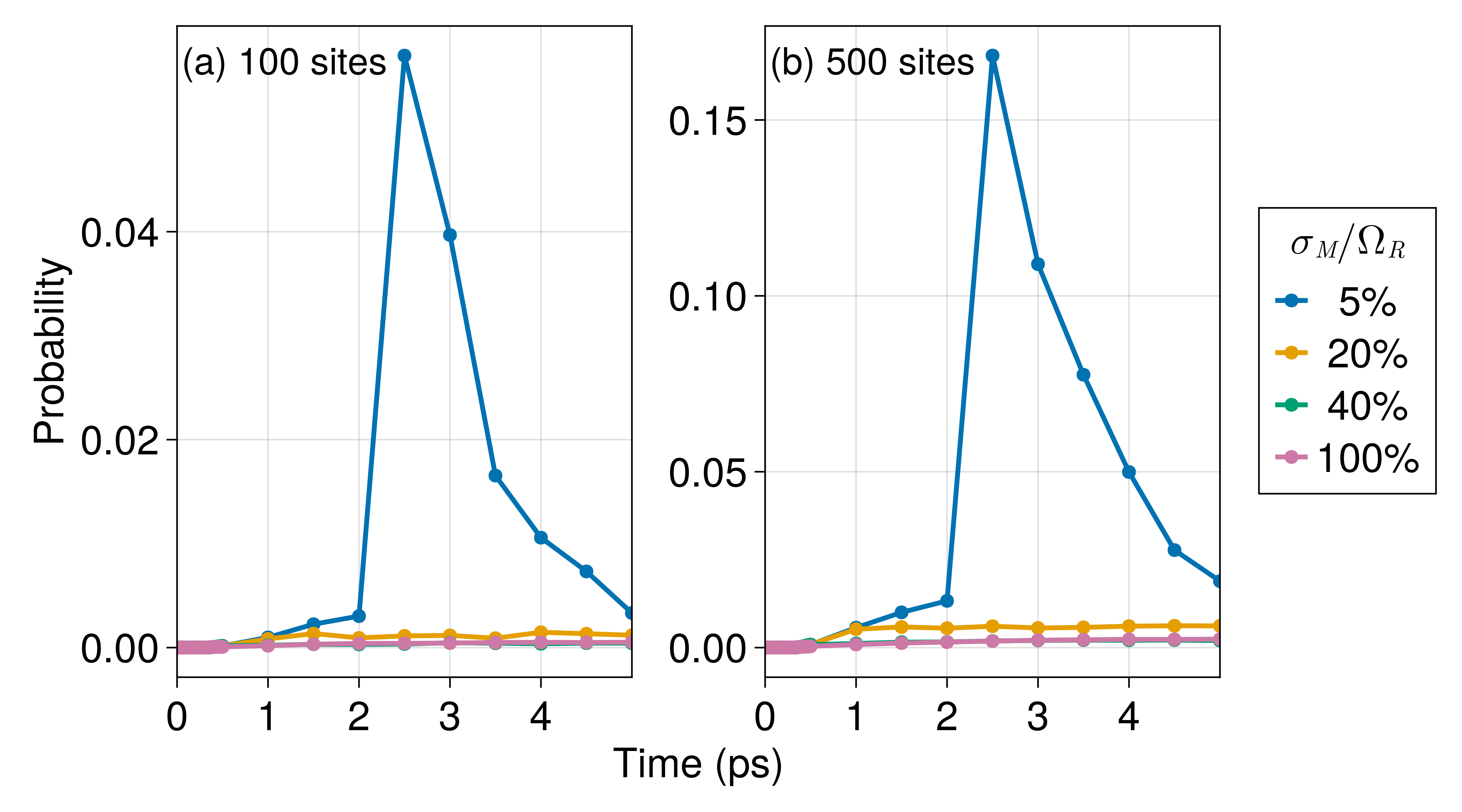}
    \caption{Probability of finding the exciton near the boundary of the wire. This value is computed by summing the excitation probability of \textbf{(a)} 100 and \textbf{(b)} 500 sites closest to the region of the wire where periodic boundary conditions are enforced ($x = 0$ and $x = 50\; \mu$m). In all cases, $\Omega_R = 0.1$ eV and the wave packet initial width and mean quasimomentum are $\sigma_x = 120$ nm and $\bar{q}_0 = 0$. Each data point is obtained from the average of 100 disorder realizations.}
    \label{fig:prob_boundary}
\end{figure}

Fig. \ref{fig:prob_boundary} shows the probability of finding an excitation on a site near the edges of the quantum wire. Since we have employed periodic boundary conditions, the exciton wave packet will go around the wire (which can be thought of more precisely as a ring) and start interfering with itself, creating finite-size artifacts. Moreover, the computation of RMSD also becomes ambiguous once the wave packet travels around. Hence, to guarantee that our results are not plagued with these artifacts, we monitor the probability of detecting the exciton at the most distant sites with respect to the initial excitation spot. From Fig. \ref{fig:prob_boundary}, we see that this probability remains small under enough disorder ($\sigma_M / \Omega_R \geq 0.2$). In the femtosecond window, even the small disorder simulation ($\sigma_M / \Omega_R = 0.05$) has a small excitation probability at the edges. These results are important to ensure that the RMSD asymptote seen in Fig. 3b (main text) is truly disorder-induced localization and not finite-size effects. 

\section{Wave packets under DET}

\begin{figure}
    \centering
    \includegraphics[width=0.8\textwidth]{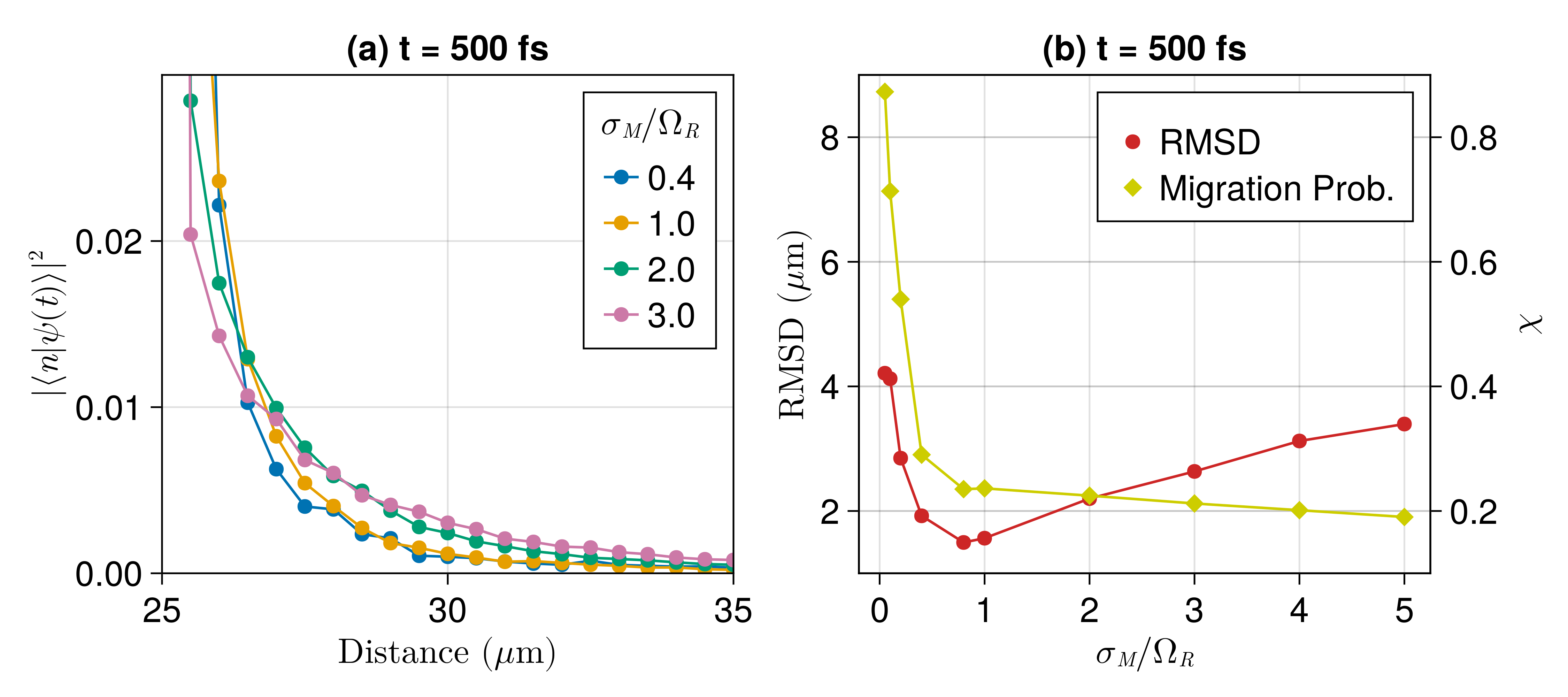}
    \caption{Wave packet properties at $t = 500$ fs: \textbf{(a)} Decay as a function of distance at different relative disorder strengths. \textbf{(b)} RMSD and migration probabilities at a fixed time step at different disorder strengths. In all cases, $\Omega_R = 0.1$ eV and $\sigma_x = 120$ nm. Other parameters are fixed as in the main text.}
    \label{fig:wvp500}
\end{figure}

\begin{figure}
    \centering
    \includegraphics[width=0.8\textwidth]{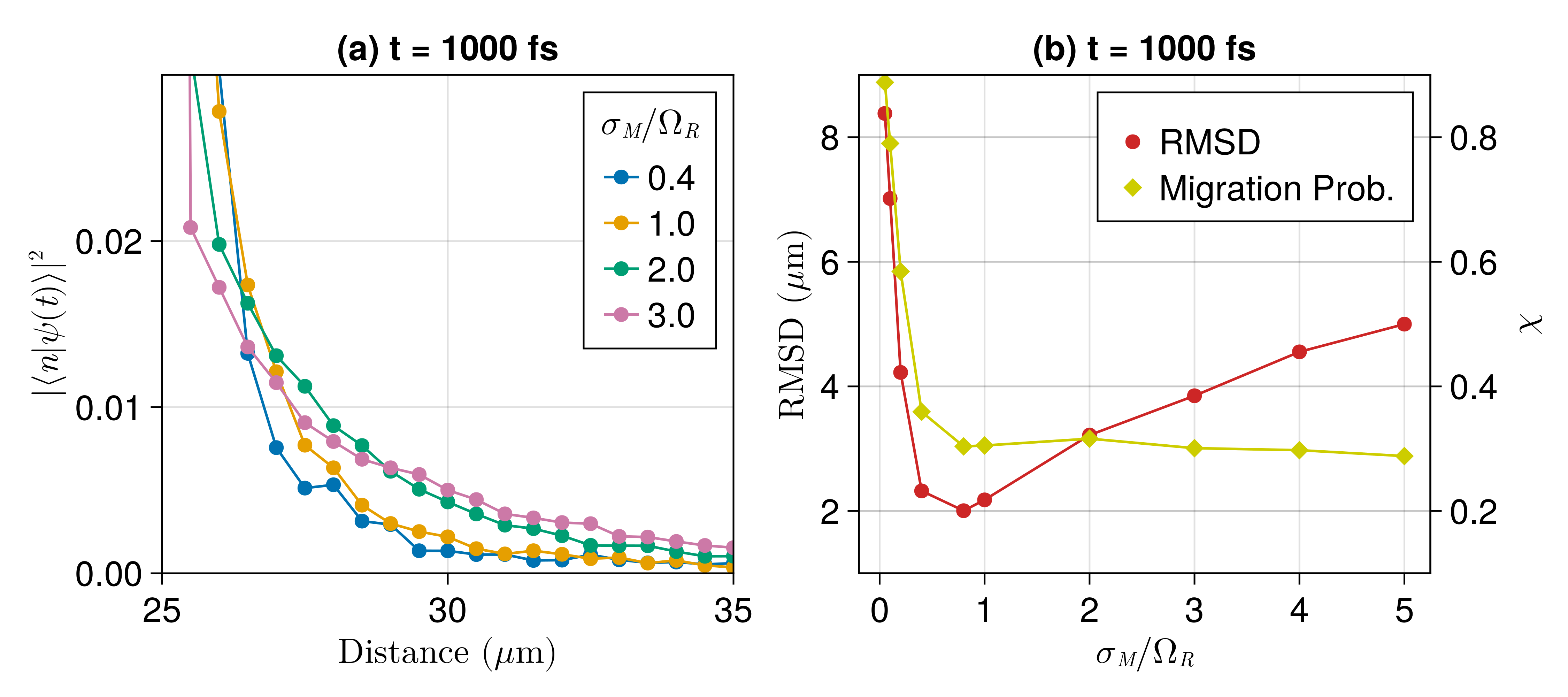}
    \caption{Wave packet properties at $t = 1000$ fs: \textbf{(a)} Decay as a function of distance at different relative disorder strengths. \textbf{(b)} RMSD and migration probabilities at a fixed time step at different disorder strengths. In all cases, $\Omega_R = 0.1$ eV and $\sigma_x = 120$ nm. Other parameters are fixed as in the main text.}
    \label{fig:wvp1000}
\end{figure}

\begin{figure}
    \centering
    \includegraphics[width=0.8\textwidth]{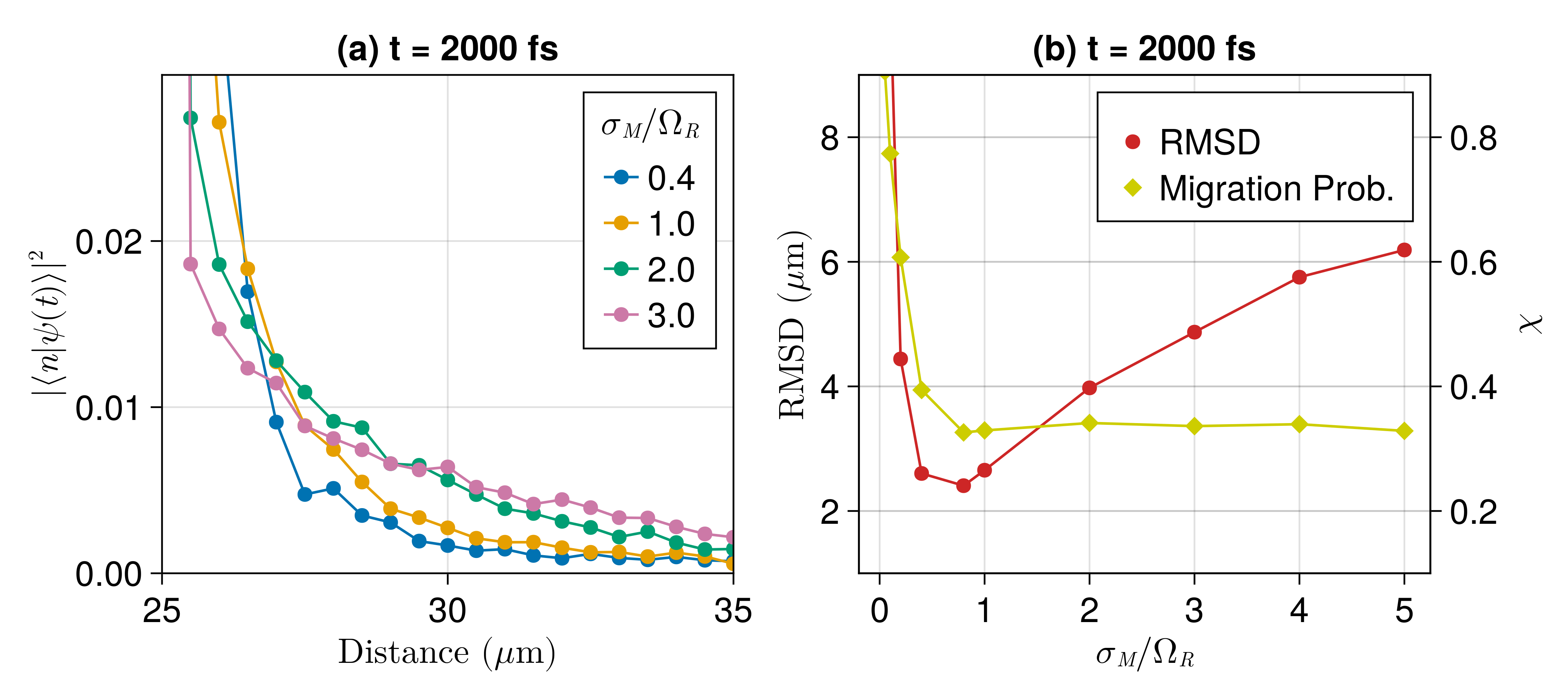}
    \caption{Wave packet properties at $t = 2000$ fs: \textbf{(a)} Decay as a function of distance at different relative disorder strengths. \textbf{(b)} RMSD and migration probabilities at a fixed time step at different disorder strengths. In all cases, $\Omega_R = 0.1$ eV and $\sigma_x = 120$ nm. Other parameters are fixed as in the main text.}
    \label{fig:wvp2000}
\end{figure}

In Figs. \ref{fig:wvp500}, \ref{fig:wvp1000}, and \ref{fig:wvp2000} we show snapshots of the exciton wave packet at different time steps where we can see the different in the wave packet shape under disorder enhanced transport (DET). In Figs. \ref{fig:wvp500}-\ref{fig:wvp2000}(a) we see that under very strong disorder ($\sigma_M/\Omega_R > 1$) the wave packet decays more quickly in space, but has a more delocalized tail. This decolalization is captured in the RMSD but not in the migration probability ($\chi$) as seen in \ref{fig:wvp500}-\ref{fig:wvp2000}(b). 

`

\clearpage
\section{Propagation Profiles}

\begin{figure}
    \centering
    \includegraphics[width=0.8\textwidth]{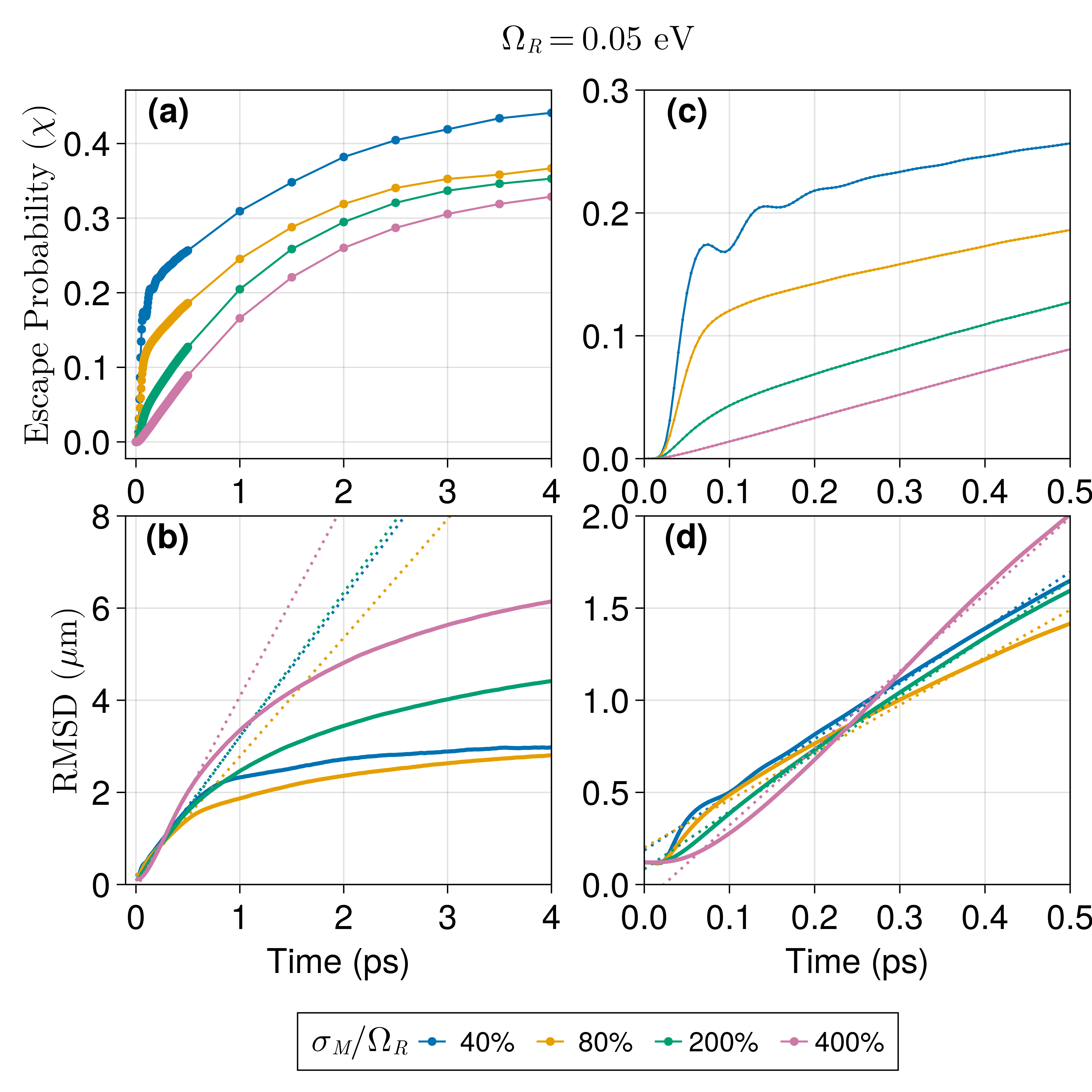}
    \caption{Propagation of exciton wave packets at different disorder strengths measured by \textbf{(a)} migration probability (Eq. \ref{SI_escape}) and \textbf{(b)} RMSD (Eq. \ref{SI_x2}). In all cases, $\sigma_x = 120$ nm and $\Omega_R = 0.05$ eV. Results are averages of 100 disordered realizations. The dotted lines in \textbf{(b)} are linear fits on the early propagation ($\leq$ 500 fs) from which slopes are used to measure the initial ballistic velocity ($v_0$). Panels \textbf{(c)} and \textbf{(d)} are zoomed in versions of \textbf{(a)} and \textbf{(b)}, respectively.}
    \label{R0p05}
\end{figure}

\begin{figure}
    \centering
    \includegraphics[width=0.8\textwidth]{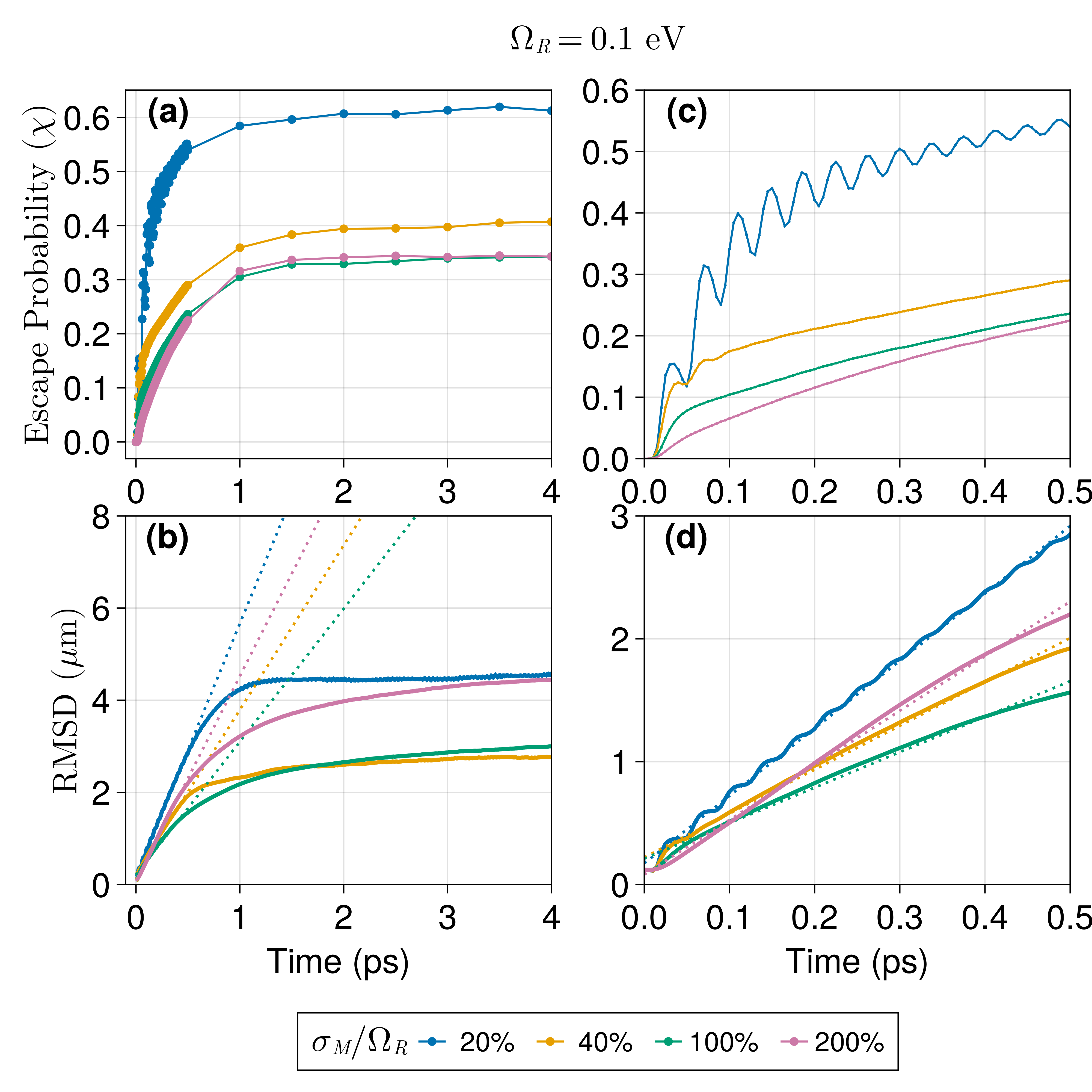}
    \caption{Propagation of exciton wave packets at different disorder strengths measured by \textbf{(a)} migration probability (Eq. \ref{SI_escape}) and \textbf{(b)} RMSD (Eq. \ref{SI_x2}). In all cases, $\sigma_x = 120$ nm and $\Omega_R = 0.1$ eV. Results are averages of 100 disordered realizations. The dotted lines in \textbf{(b)} are linear fits on the early propagation ($\leq$ 500 fs) from which slopes are used to measure the initial ballistic velocity ($v_0$). Panels \textbf{(c)} and \textbf{(d)} are zoomed in versions of \textbf{(a)} and \textbf{(b)}, respectively.}
    \label{R0p1}
\end{figure}

\begin{figure}
    \centering
    \includegraphics[width=0.8\textwidth]{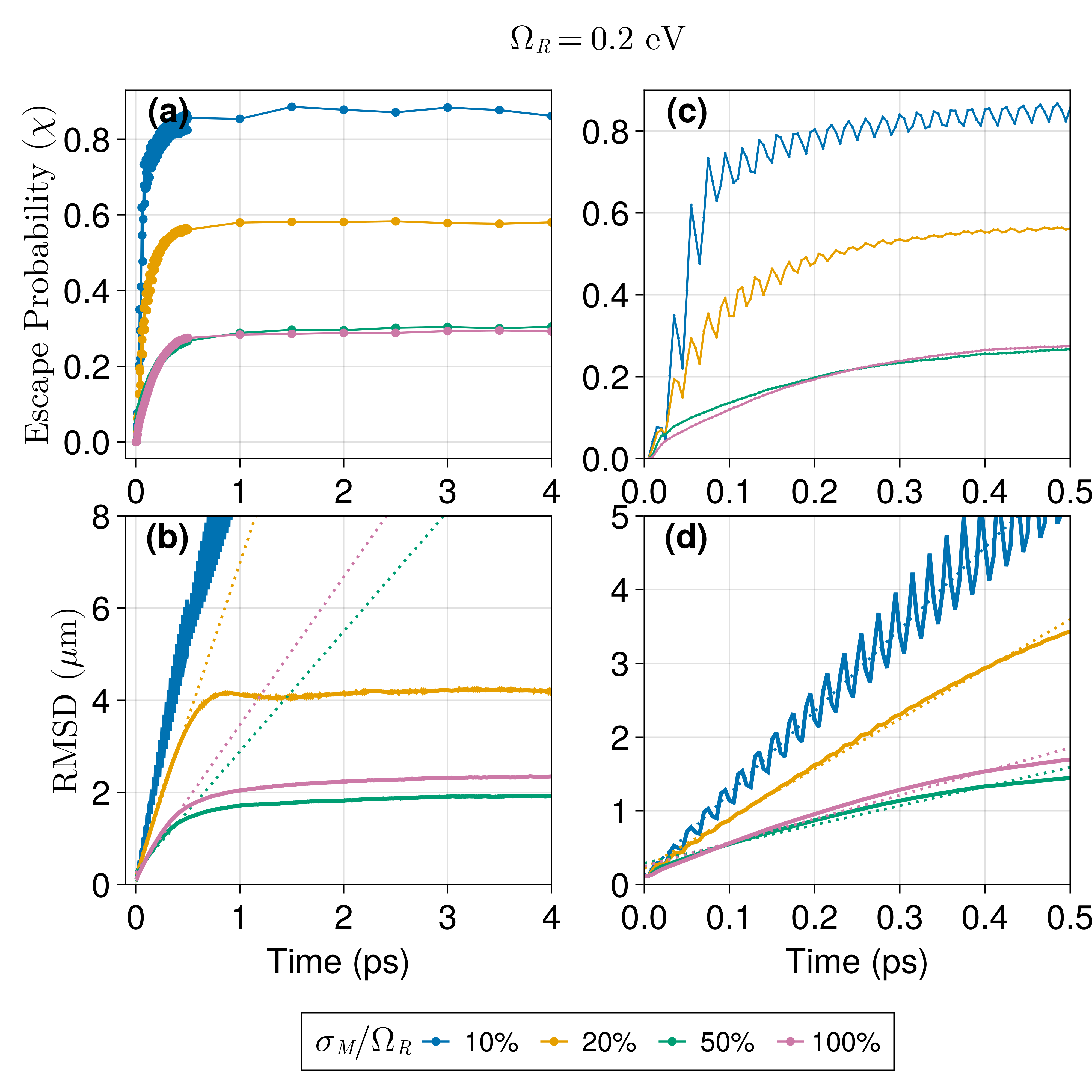}
    \caption{Propagation of exciton wave packets at different disorder strengths measured by \textbf{(a)} migration probability (Eq. \ref{SI_escape}) and \textbf{(b)} RMSD (Eq. \ref{SI_x2}). In all cases, $\sigma_x = 120$ nm and $\Omega_R = 0.2$ eV. Results are averages of 100 disordered realizations. The dotted lines in \textbf{(b)} are linear fits on the early propagation ($\leq$ 500 fs) from which slopes are used to measure the initial ballistic velocity ($v_0$). Panels \textbf{(c)} and \textbf{(d)} are zoomed in versions of \textbf{(a)} and \textbf{(b)}, respectively.}
    \label{R0p2}
\end{figure}

\begin{figure}
    \centering
    \includegraphics[width=0.8\textwidth]{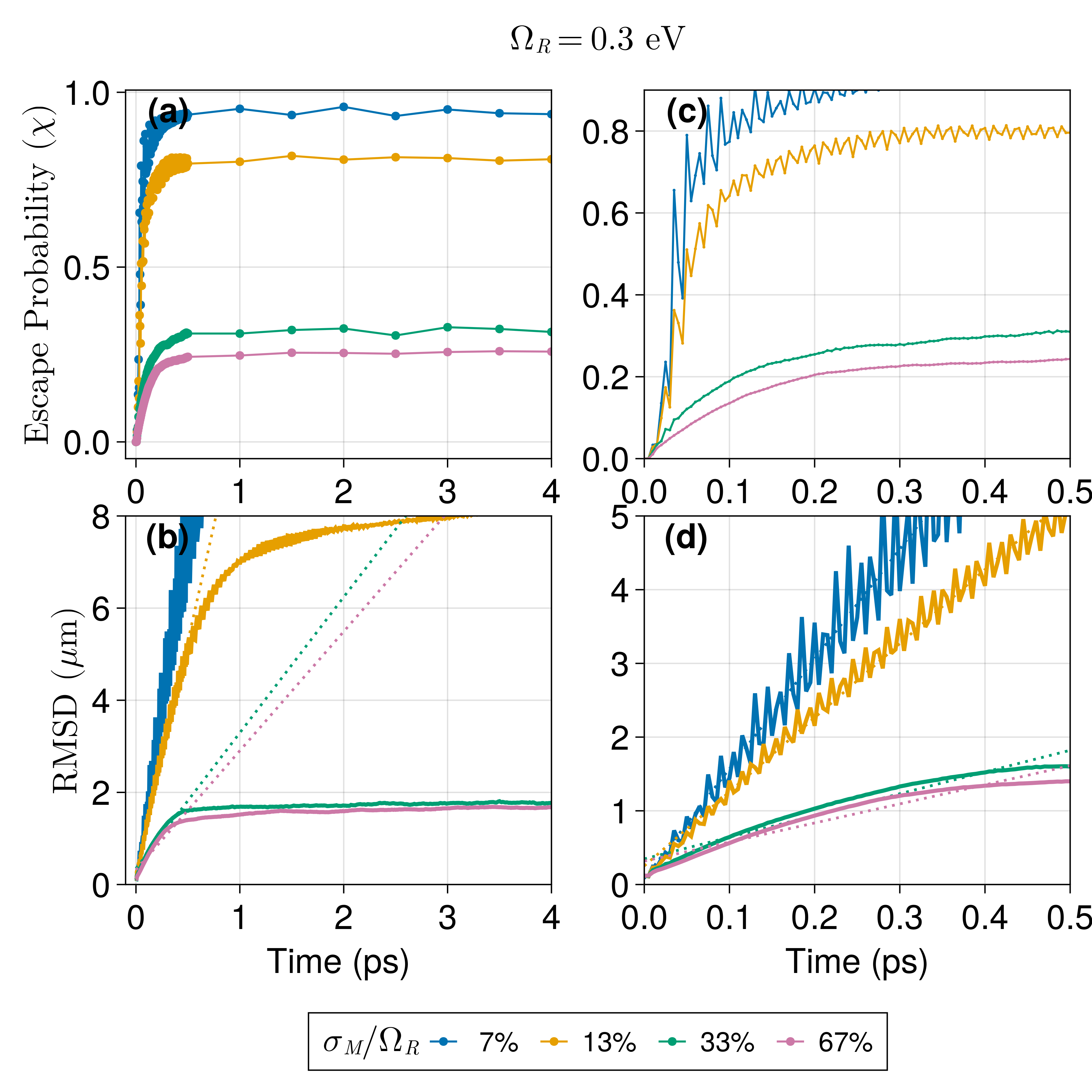}
    \caption{Propagation of exciton wave packets at different disorder strengths measured by \textbf{(a)} migration probability (Eq. \ref{SI_escape}) and \textbf{(b)} RMSD (Eq. \ref{SI_x2}). In all cases, $\sigma_x = 120$ nm and $\Omega_R = 0.3$ eV. Results are averages of 100 disordered realizations. The dotted lines in \textbf{(b)} are linear fits on the early propagation ($\leq$ 500 fs) from which slopes are used to measure the initial ballistic velocity ($v_0$). Panels \textbf{(c)} and \textbf{(d)} are zoomed in versions of \textbf{(a)} and \textbf{(b)}, respectively.}
    \label{R0p3}
\end{figure}

Figs. \ref{R0p05}-\ref{R0p3} shows the time evolution of the RMSD and migration probability ($\chi$) defined in the main text and repeated below for reference
\begin{align}
    \text{RMSD}(t) &= \left[\frac{1}{P_\text{M}(t)}\sum_n^{N_M} |\bra{n;0}\ket{\psi(t)}|^2(x_n - x_0)^2\right]^{1/2}\,, \label{SI_x2}\\
    P_\text{M}(t) &= \sum_n^{N_M} |\bra{n;0}\ket{\psi(t)}|^2 \,. \\
    \chi(t) &= 1 - \frac{1}{P_\text{M}(t)}\sum_{n = n_\text{min}}^{n_\text{max}} |\bra{n;0}\ket{\psi(t)}|^2 \label{SI_escape} \,.
\end{align}
It can be seen that the time to reach a steady state (constant RMSD) decreases as the Rabi splitting is increased. 

\clearpage
\section{Linear Fit Range}

\begin{figure*}
    \centering
    \includegraphics[width=0.8\textwidth]{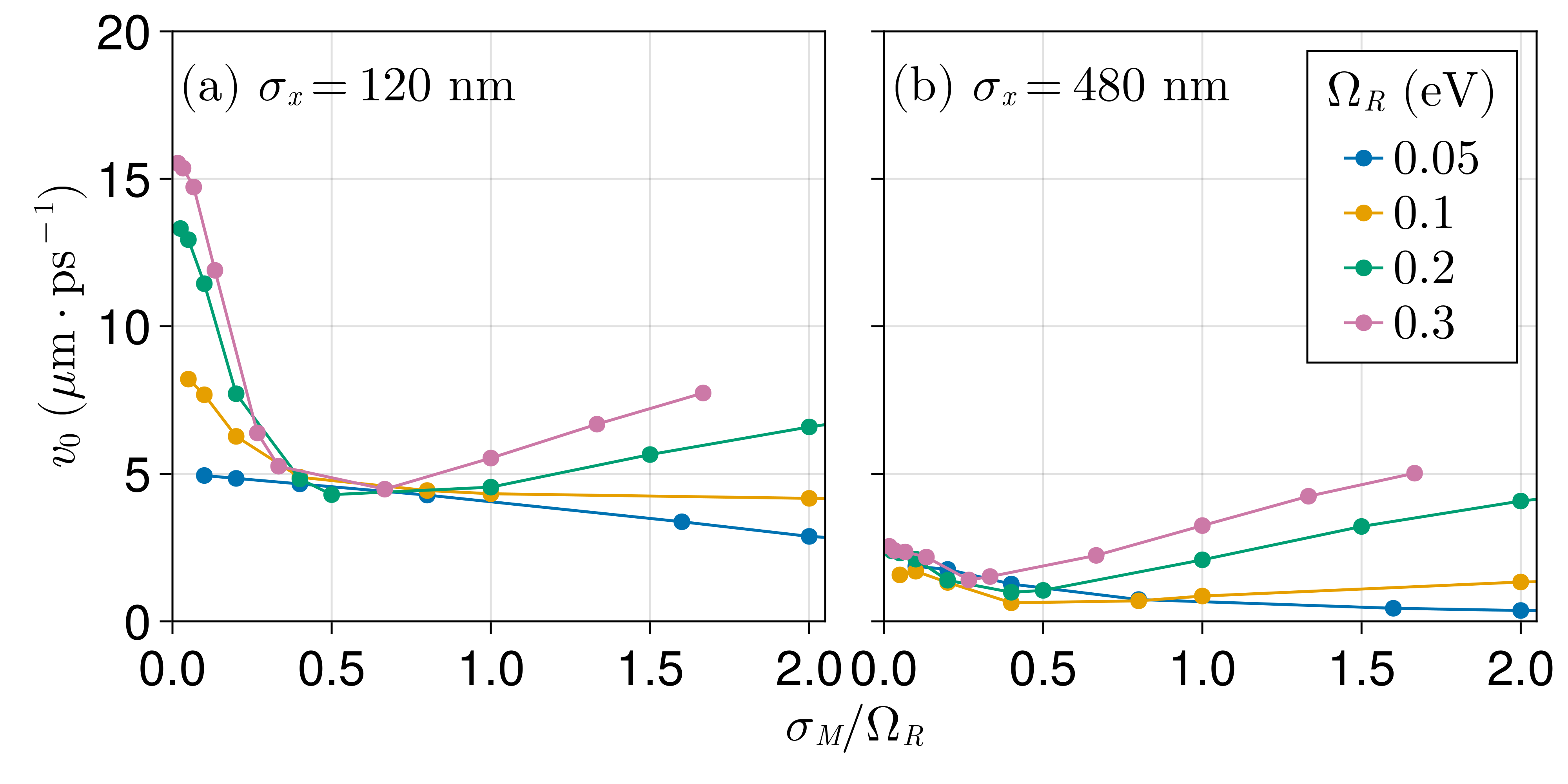}
    \caption{Initial spread velocity ($v_0$) for various systems at different relative disorder strength and Rabi splitting values for wave packets with a \textbf{(a)} narrow and \textbf{b} broad initial spread values ($\sigma_x$). Each point is the average of 100 disordered realizations. $v_0$ was computed as the slope of a linear fit of RMSD values in the initial 100 fs of simulation (see Figure 3) in the main text.}
    \label{fig:v0_0p1}
\end{figure*}

\begin{figure*}
    \centering
    \includegraphics[width=0.8\textwidth]{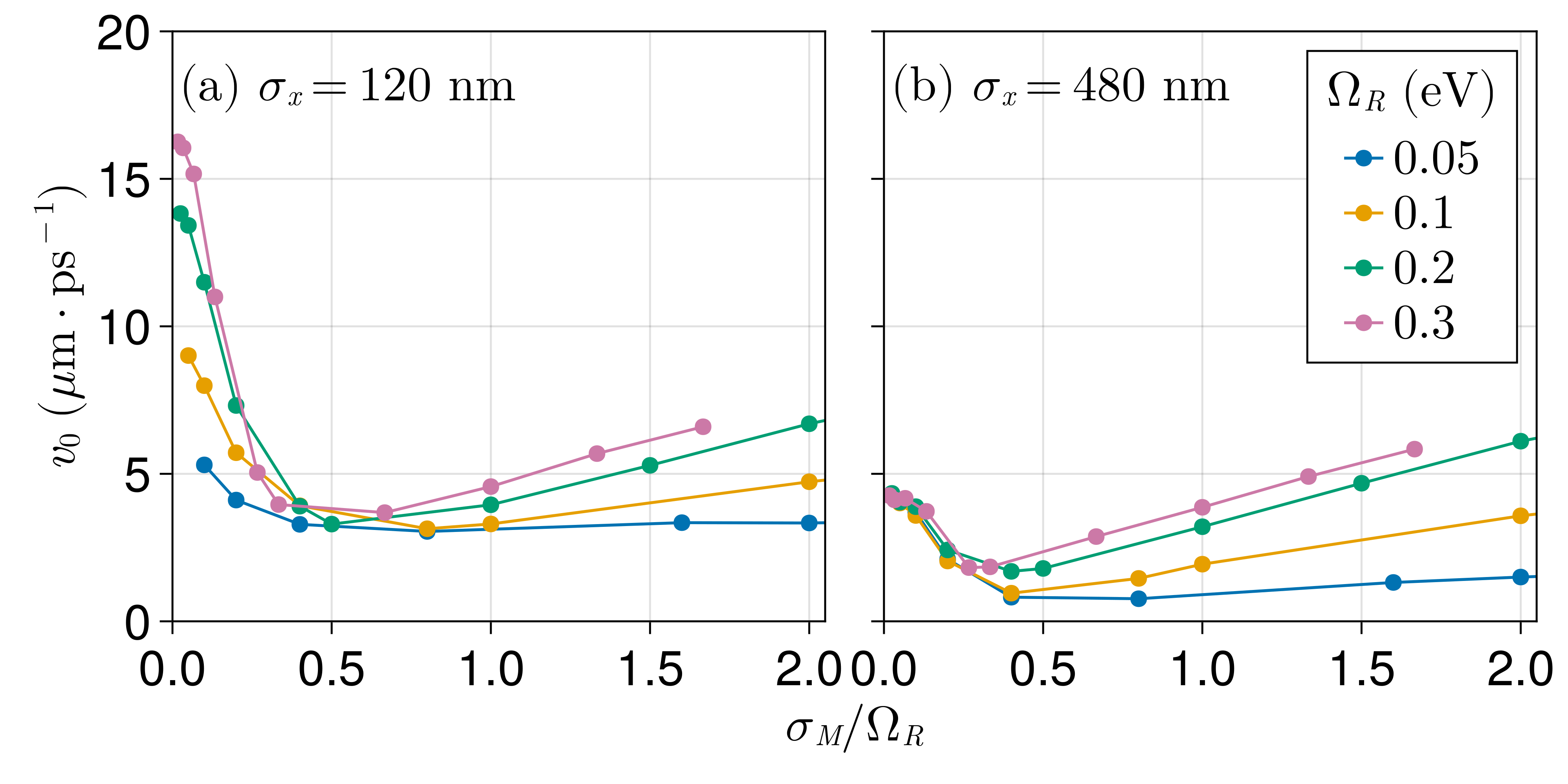}
    \caption{Initial spread velocity ($v_0$) for various systems at different relative disorder strength and Rabi splitting values for wave packets with a \textbf{(a)} narrow and \textbf{b} broad initial spread values ($\sigma_x$). Each point is the average of 100 disordered realizations. $v_0$ was computed as the slope of a linear fit of RMSD values in the initial 300 fs of simulation (see Figure 3) in the main text.}
    \label{fig:v0_0p3}
\end{figure*}

\begin{figure*}
    \centering
    \includegraphics[width=0.8\textwidth]{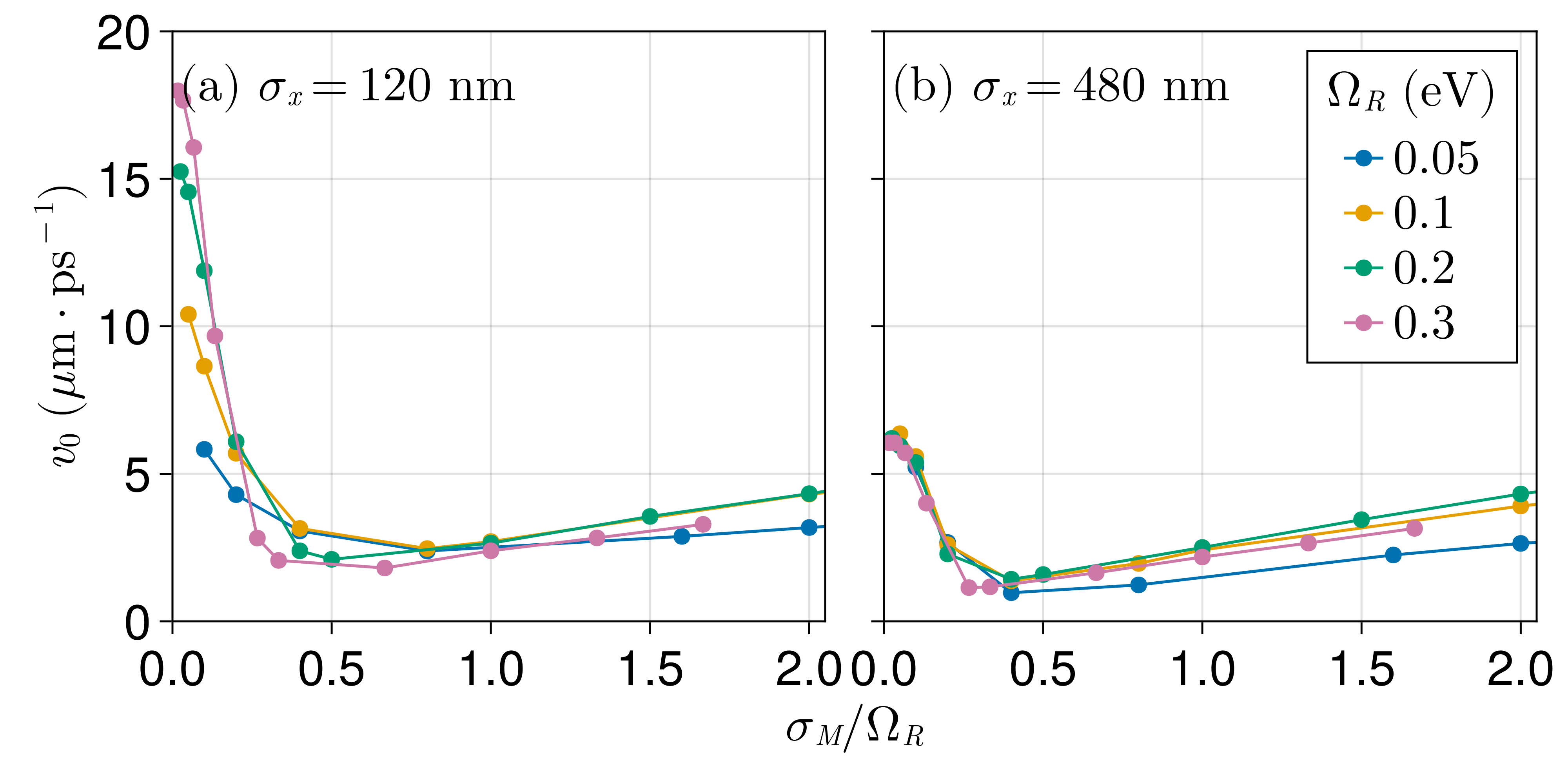}
    \caption{Initial spread velocity ($v_0$) for various systems at different relative disorder strength and Rabi splitting values for wave packets with a \textbf{(a)} narrow and \textbf{b} broad initial spread values ($\sigma_x$). Each point is the average of 100 disordered realizations. $v_0$ was computed as the slope of a linear fit of RMSD values in the initial 700 fs of simulation (see Figure 3) in the main text.}
    \label{fig:v0_0p7}
\end{figure*}

In the main text, we have used the first 500 fs of our simulations to compute the ballistic velocity ($v_0$) using a linear fit. Our criterion to choose an interval for a linear fit was 1) long enough periods such that complicated initial features are averaged out, and 2) short enough to prevent the consideration of time steps where Anderson localization starts to appear (around 1 ps). Figs. \ref{fig:v0_0p1}, \ref{fig:v0_0p3}, and \ref{fig:v0_0p7} show analogs of Figure 4 (main text) using different time intervals for the linear fit (0.1, 0.3, and 0.7 ps, respectively). The choice of time interval is not strictly fundamental for a qualitative analysis as we see an overall agreement for all time intervals. Importantly, The $\sigma_x$ dependence discussed in the text is maintained in all cases. 

\clearpage
\section{$v_0$ for detuned cavities}

\begin{figure}
    \centering
    \includegraphics[width=\textwidth]{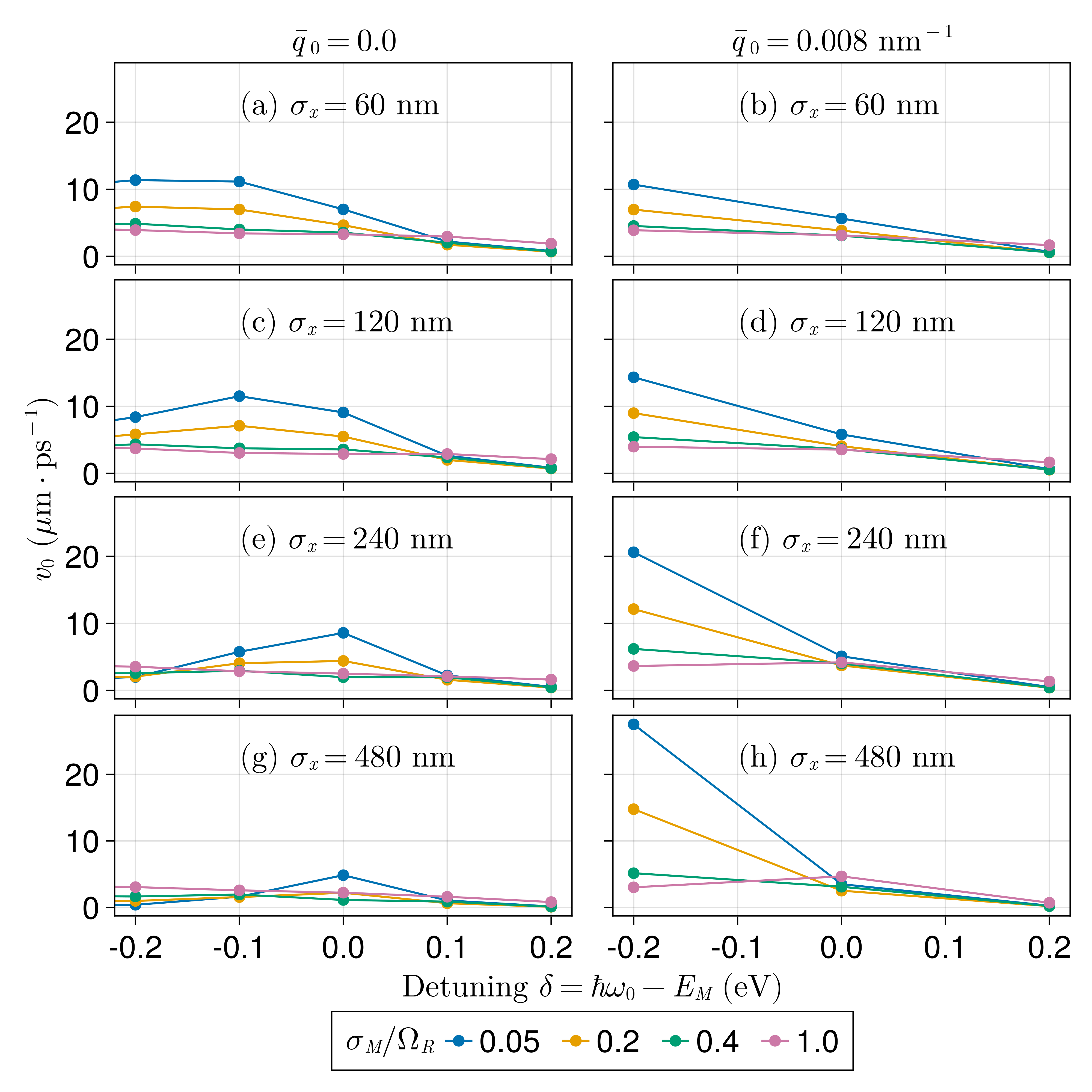}
    \caption{Ballistic velocity ($v_0$) as a function of the cavity detuning ($\delta$) for several disorder strengths ($\sigma_M / \Omega_R$). Left column panels (a, c, e, g) show results for excitons with zero quasi-momentum ($\bar{q}_0 = 0$) whereas in the right column panels excitons have $\bar{q}_0 = 0.008$ nm$^{-1}$. In each panel, a different initial spread is used ($\sigma_x$).}
    \label{fig:v0detun}
\end{figure}

Figure \ref{fig:v0detun} shows $v_0$ (see Figure 3b on the main text for more details about how $v_0$ is obtained) as a function of detuning for several initial excitons spread values ($\sigma_x$). In particular, the left column panels (a, c, e, g) display results using a zero average initial momentum ($\bar{q}_0 = 0$) while the right column (b, d, f, h) shows results for a non-zero quasi-momentum ($\bar{q}_0 = 0.008$ nm$^{-1}$). Results presented in this figure numerically verify the predictions from the effective group velocity model (see Figure 8 in the main text). Namely, we see that when $\bar{q}_0 = 0$:
\begin{enumerate}
    \item Narrow wave packets have higher mobility.
    \item Blueshifted cavities have always slower excitons than resonant ones.
    \item Redshifted cavities have more mobility excitons if the wave packets are narrow enough.
\end{enumerate}
For $\bar{q}_0 = 0.008$ nm$^{-1}$ (note that this is near the $v^\text{eff}_{\alpha q}$ peak for the redshifted cavity in Figure 8), we verify the following predictions:
\begin{enumerate}
    \item Redshifted cavities have faster excitons than resonant ones.
    \item Broader wave packets are more mobile than narrow ones.
    \item Blueshifted cavities have always slower excitons than resonant ones.
\end{enumerate}

\clearpage

\begin{figure}
    \centering
    \includegraphics[width=\textwidth]{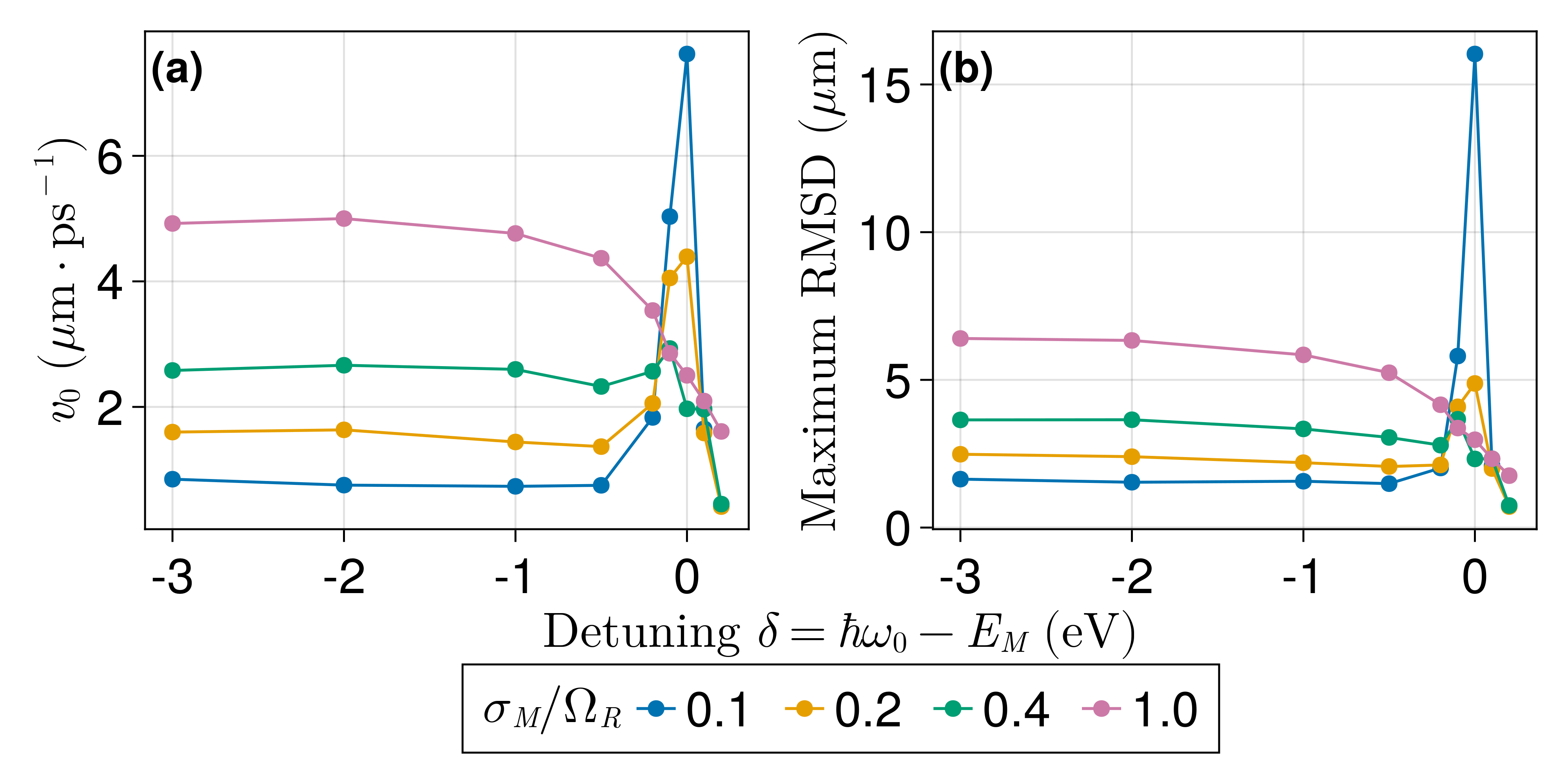}
    \caption{Disorder-dependent detuning effects on coherent exciton transport measured by \textbf{(a)} the ballistic velocity ($v_0$), and \textbf{(b)} the maximum RMSD over 5 ps. In all cases $\sigma_x = 240$ nm and $\Omega_R = 0.1$ eV. Each data point corresponds to an average over 100 disorder realizations.}
    \label{fig:v0detunextra}
\end{figure}

Figure \ref{fig:v0detunextra} shows how $v_0$ and the maximum RMSD depend on the cavity detuning. This is similar to Figure 7 of the main text, but with more redshifted values to explore whether an extremum exists in this direction. While we do not observe a peak, we see that both $v_0$ and the maximum RMSD reach an asymptotic behavior, i.e. further changes in the cavity detuning (towards negative values) do not change the overall exciton mobility. We highlight that, for very redshifted cavities, the description of the photonic degrees of freedom becomes poorer as higher energy bands ($n_y,\; n_z > 1$) become more relevant. Therefore, these results represent the theoretical limit in a single-band cavity system.

\clearpage
\section{Strong Coupling Signatures}

\par
\begin{figure}
    \centering
    \includegraphics[width=\textwidth]{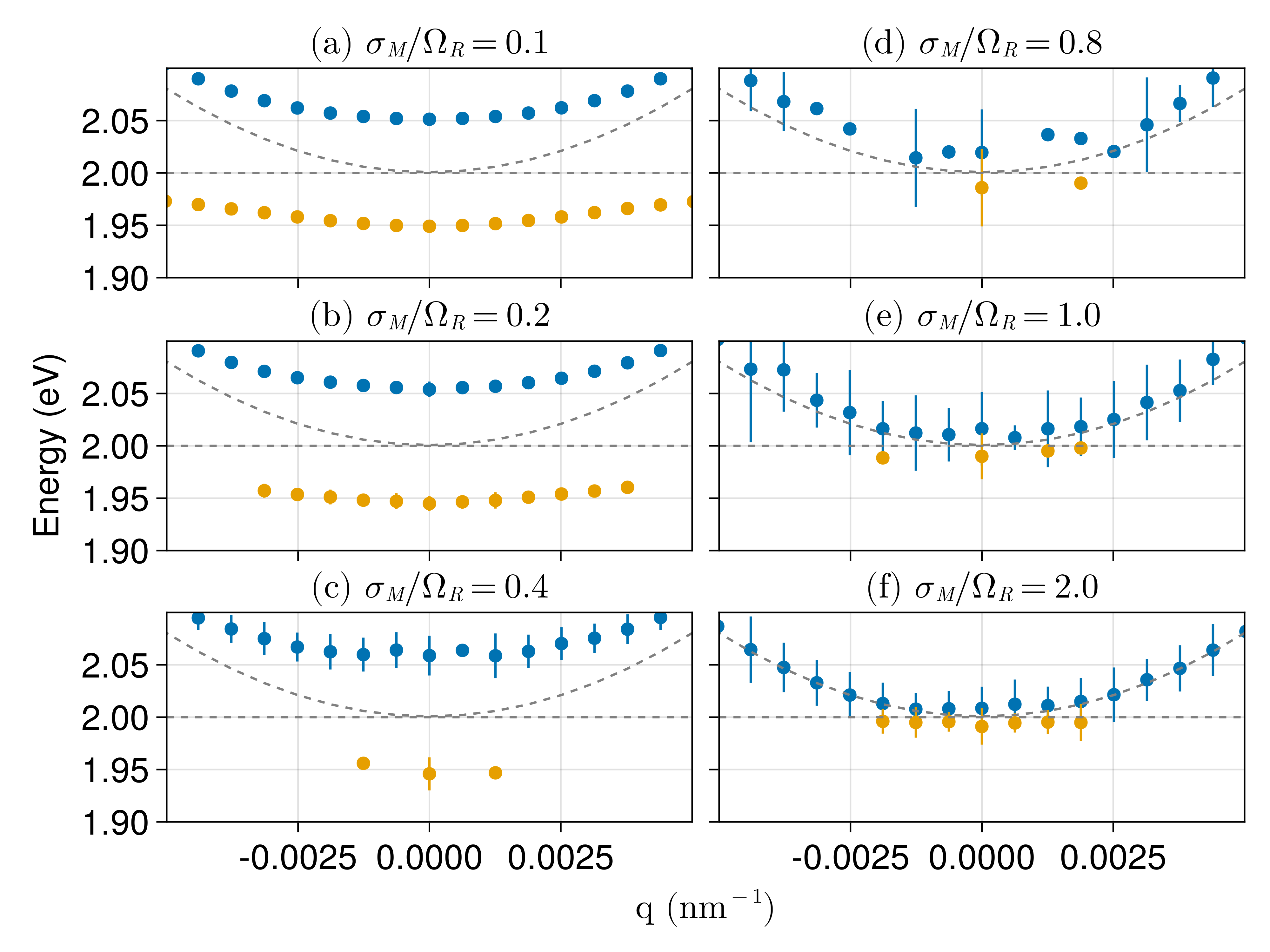}
    \caption{Average energy vs photon $q$ with maximum probability for bright modes (total photon content $> 10\%$) of systems under increasing disorder. Each point represent the mean energy obtained from the eigenstates with more than $10\%$ photonic content in 25 realizations of the light-matter Hamiltonian with the specified $\sigma_M/\Omega_R$ and $\Omega_R$ fixed at 0.1 eV. For the sake of simplicity, we show results for systems containing $N_M = 1000$ molecules and 401 photonic modes. The graphs show the gradual closing of the polariton gap (here represented by $E_{\text{UP},q} - E_{\text{LP},q}$ with $q=0$) as $\sigma_M/\Omega_R$ increases. Finer features (e.g., the number of points in the LP and UP belonging to specific intervals of $q$) are unlikely to be converged due to the small system size and the number of realizations we employed. However, this lack of convergence is irrelevant to our analysis of the polariton gap.}
    \label{fig:dispersion}
\end{figure}

 Figure \ref{fig:dispersion} shows mean energy (and fluctuations) vs maximum value of photonic $q$ for the eigenstates with more than $10\%$ photonic content obtained from 25 realizations of a system of 1000 dipoles coupled to 401 photonic modes. As expected, at small and moderate disorder ($\sigma_M/\Omega_R \leq 0.4$), two distinct curves separated by approximately $\Omega_R$ are observed, corresponding to LP (yellow) and UP (blue) branches of the quasiparticles emergent from strong light-matter coupling. Figure S15 also shows that further increase of disorder shrinks the gap between these two branches, and when the relative disorder $\sigma_M / \Omega_R$ approaches 1, it is no longer possible in general to discern the two branches at the resonance wave vector ($q = 0$). That is, the characteristic splitting associated with the strong coupling regime is no longer observable. We ascribe no meaning to the irregularities and breakdown of inversion symmetry at large and moderate disorder, as these features likely reflect the smallness of the system size and the limited number of disorder realizations we employed to identify the breakdown of strong coupling (as defined by the gap between LP and UP at the resonant wave vector $q=0$).

\begin{figure}
    \centering
    \includegraphics[width=\textwidth]{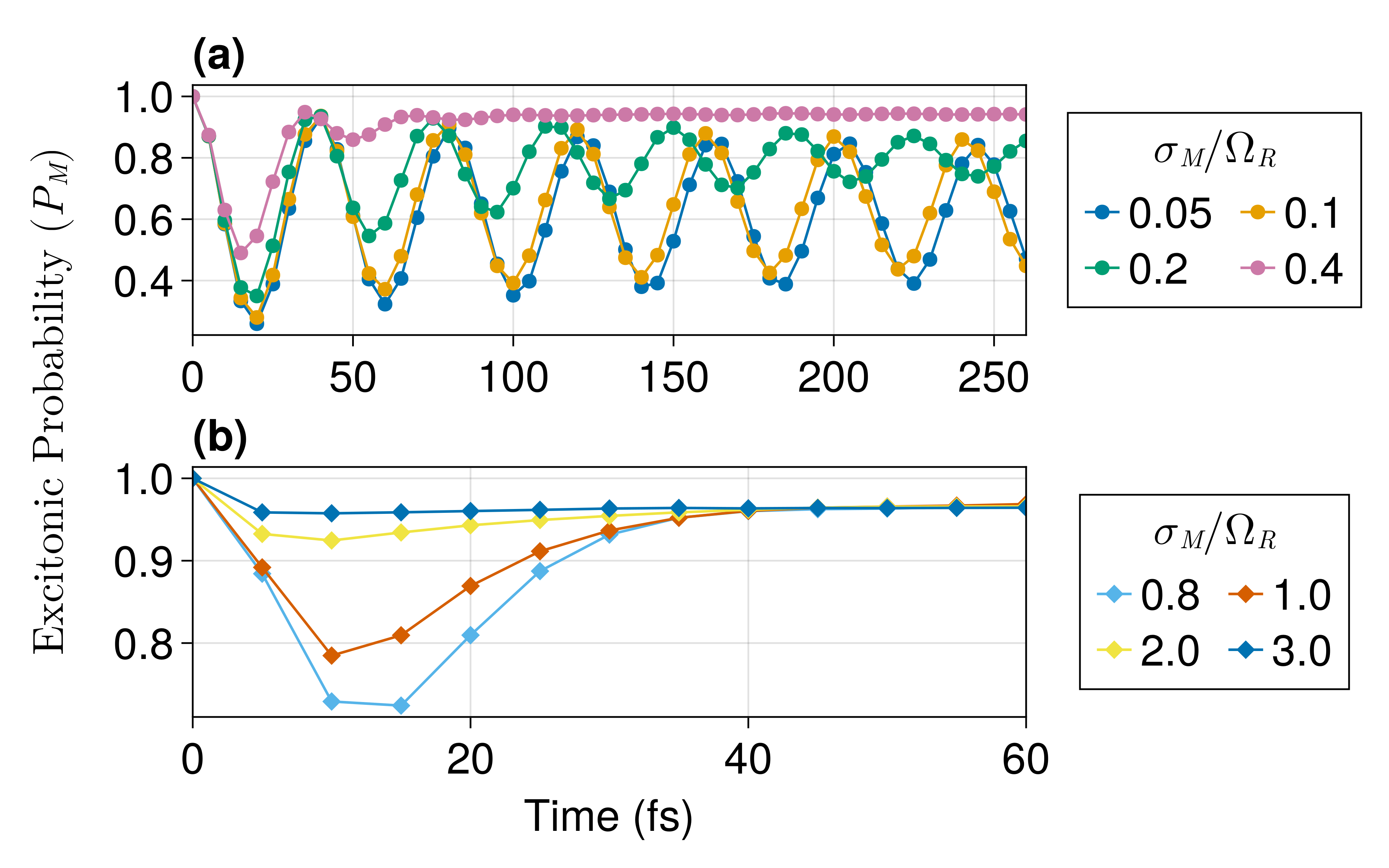}
    \caption{Demonstration of the effect of disorder on Rabi oscillations. The excitonic probability is the summed probability that any site is in its excited state. In all cases the Rabi splitting is fixed at 0.1 eV and each point is the average of 100 realizations.}
    \label{fig:rabiosc}
\end{figure}

Figure \ref{fig:rabiosc} presents the change in the exciton content of the wave packet over time where Rabi oscillations can be observed. Under weak disorder \textbf{(a)} clear oscillations with a frequency of approximately 41 fs$^{-1}$ (which corresponds to the Rabi splitting of 0.1 eV used here) can be verified. Under stronger values of disorder, shown in Figure \ref{fig:rabiosc}b, we see that Rabi oscillations are strongly suppressed. When $\sigma_M / \Omega_R$ is close to one, is no longer possible to identify any oscillation.

The results presented in Figures \ref{fig:dispersion} and \ref{fig:rabiosc} are typically employed in criteria for the strong coupling regime. Despite the absence of dissipation in our model, we see that strong static disorder is capable of removing the typical signatures of the strong coupling regime. Hence, one may consider that when the static disorder is comparable to the Rabi splitting (that is, the expected collective coupling value) a different regime emerges. Nonetheless, our model still predicts coherent transport, therefore we consider it to be distinct from the weak coupling situation where non-interacting molecules would not exchange energy.

\end{document}